\pdfoutput=1
\documentclass{article}

\usepackage{graphicx, fancyhdr, amssymb, amsmath, amsthm, 
  epsfig, mathrsfs, verbatim}

 \numberwithin{equation}{section}
\usepackage[section]{placeins}
\usepackage{natbib}
\usepackage{graphicx,graphics, fancyhdr,fancyvrb, amssymb,hyperref, amsmath,amsthm,mathrsfs , bm,algorithmicx,algorithm,algpseudocode,relsize}
\usepackage{footmisc}

\usepackage[margin=2cm]{geometry}
\usepackage[none]{hyphenat}
\usepackage{footnote,tablefootnote}
\usepackage{multirow}

\usepackage[compact]{titlesec}
\titlespacing{\section}{0pt}{*0}{*0}
\titlespacing{\subsection}{0pt}{*0}{*0}
\titlespacing{\subsubsection}{0pt}{*0}{*0}

\makesavenoteenv{tabular}
 \newcommand{\sgn}{\operatorname{sgn}}
\newcommand{\argmin}{\operatornamewithlimits{argmin}}
\newcommand{\utwi}[1]{\mbox{\boldmath $ #1$}}
\DeclareMathOperator{\Tr}{Tr}
\usepackage{array}

\newcounter{exno}

\makeatletter
\newcommand*\rel@kern[1]{\kern#1\dimexpr\macc@kerna}
\newcommand*\widebar[1]{%
  \begingroup
  \def\mathaccent##1##2{%
    \rel@kern{0.8}%
    \overline{\rel@kern{-0.8}\macc@nucleus\rel@kern{0.2}}%
    \rel@kern{-0.2}%
  }%
  \macc@depth\@ne
  \let\math@bgroup\@empty \let\math@egroup\macc@set@skewchar
  \mathsurround\z@ \frozen@everymath{\mathgroup\macc@group\relax}%
  \macc@set@skewchar\relax
  \let\mathaccentV\macc@nested@a
  \macc@nested@a\relax111{#1}%
  \endgroup
}
\makeatother

\newcommand{\Y}{{\utwi{Y}}}
\newcommand{\X}{{\utwi{X}}}
\newcommand{\B}{{\utwi{B}}}
\newcommand{\Z}{{\utwi{Z}}}
\newcommand{\PhiB}{{\utwi{\Phi}}}
\newcommand{\betaB}{{\utwi{\beta}}}
\title{VARX-L: Structured Regularization for Large Vector Autoregressions with Exogenous Variables  
}\author{William B. Nicholson\footnote{
Corresponding Author,
PhD Candidate,  
Department of Statistical Science,
Cornell University,
301 Malott Hall,
Ithaca, NY 14853
(E-mail: \href{mailto:wbn8@cornell.edu}{wbn8@cornell.edu}; Webpage: \url{http://www.wbnicholson.com})},
  David S. Matteson\footnote{ 
Assistant Professor, 
Department of Statistical Science and Department of Social Statistics,
Cornell University,
1196 Comstock Hall,
Ithaca, NY 14853,
(E-mail: \href{mailto:matteson@cornell.edu}{matteson@cornell.edu}; Webpage: \url{http://www.stat.cornell.edu/~matteson/})},
and Jacob Bien\footnote{
Assistant Professor, 
Department of Biological Statistics and Computational Biology and Department of Statistical Science,
Cornell University,
1178 Comstock Hall,
Ithaca, NY 14853
(E-mail: \href{mailto:jbien@cornell.edu}{jbien@cornell.edu}; Webpage: \url{http://faculty.bscb.cornell.edu/\textasciitilde bien/})}
}
\date{\today}
\usepackage[singlespacing]{setspace}

\begin{document}
\maketitle
\begin{abstract}
The vector autoregression (VAR)
has long proven to be an effective method for modeling the joint dynamics of macroeconomic time series as well as forecasting.
A major shortcoming of the VAR that has hindered its applicability is its heavy parameterization: the parameter space grows quadratically with the number of series included, quickly exhausting the available degrees of freedom.  Consequently, forecasting using VARs is intractable for low-frequency, high-dimensional macroeconomic data.  However, empirical evidence suggests that VARs that incorporate more component series tend to result in more accurate forecasts.  Conventional methods that allow for the estimation of large VARs either tend to require \emph{ad hoc} subjective specifications or are computationally infeasible. Moreover, as global economies become more intricately intertwined, there has been substantial interest in incorporating the impact of stochastic, unmodeled  \emph{exogenous} variables.  Vector autoregression with exogenous variables (VARX) extends the VAR to allow for the inclusion of unmodeled variables, but it similarly faces dimensionality challenges.       

We introduce the VARX-L framework, a structured family of VARX models, and provide methodology that allows for both efficient estimation and accurate forecasting in high-dimensional analysis. VARX-L adapts several prominent scalar regression regularization techniques to a vector time series context in order to greatly reduce the parameter space of VAR and VARX models.
We also highlight a compelling extension that allows for shrinking toward reference models, such as a vector random walk. We demonstrate the efficacy of VARX-L in both low- and high-dimensional macroeconomic forecasting applications and simulated data examples.  Our methodology is easily reproducible in a publicly available {\tt R} package.   

\end{abstract}   
\smallskip
\noindent \textbf{Keywords: }{Big Data, Forecasting, Group Lasso, Macroeconometrics, Time Series}
\newpage
\section{Introduction}
\setstretch{2}
The practice of macroeconomic forecasting was spearheaded by \cite{klein}, whose eponymous simultaneous equation system jointly forecasted the behavior of 15 annual macroeconomic indicators, including consumer expenditures, interest rates, and corporate profits.  The parameterization and identification restrictions of these models were heavily influenced by Keynesian economic theory.  As computing power increased, such models became larger and began to utilize higher frequency data.  Forecasts and simulations from these models were commonly used to inform government policymakers as to the overall state of the economy and to influence policy decisions \citep{welfe}.  As the Klein-Goldberger model and its extensions were primarily motivated by Keynesian economic theory, the collapse of the Bretton Woods monetary system and severe oil price shocks led to widespread forecasting failure in the 1970s \citep{diebold}.  At this time, the vector autoregression (VAR), popularized by \cite{sims1980}, emerged as an atheoretical forecasting technique underpinned by statistical methodology and not subject to the ebb and flow of contemporary macroeconomic theory.

Unfortunately, the VAR's flexibility can create modeling complications.  In the absence of prior information, the VAR assumes that every series interacts linearly with both its own past values as well as those of every other included series.  Such a model is known as an \emph{unrestricted} VAR.  As most economic series are low-frequency (monthly, quarterly, or annual) there is rarely enough available data to accurately forecast using large unrestricted VARs.  Such models are overparameterized, provide inaccurate forecasts, and are very sensitive to changes in economic variables.  Consequently, in such applications, the VAR's parameter space must be reduced, either in a data-driven manner or based upon the modeler's knowledge of the underlying economic system.  This model selection process has been described as ``blending data and personal beliefs according to a subjective, undocumented procedure that others cannot duplicate'' \citep{Todd}[p. 18].  

Despite their overparameterization, in many applications, large VARs can be preferable to their smaller counterparts, as small models can exclude potentially relevant variables.  Ideally, if one has no prior knowledge that a variable is irrelevant, it should be included in the model.  For example, as described in \cite{LutkDR}, modeling the Taylor Rule \citep{Taylor} requires an estimate of the ``output gap'' between real Gross Domestic Product and potential output.  The output gap is difficult to measure and can include many explanatory variables encompassing disaggregated economic measurements.  Moreover, recent work by \cite{Ibarra} and \cite{Hendry} have shown that incorporating disaggregated series improves upon the forecasts of macroeconomic aggregates such as the Consumer Price Index.  Hence, in these scenarios, to properly utilize all relevant economic information, a large vector autoregression with a coherent variable selection procedure is required.

Shortly after the VAR's inception, efforts were made to develop a systematic approach to reduce its parameterization.  Early attempts, such as \cite{Litterman1979}, centered upon a Bayesian approach underpinned by contemporary macroeconomic theory.  In applying a Bayesian VAR with a Gaussian prior (analogous to ridge regression), specific priors were formulated based upon stylized facts regarding US macroeconomic data.  For example, the popular \emph{Minnesota prior} incorporates the prevailing belief that macroeconomic variables can be reasonably modeled by a univariate random walk via shrinking estimated models toward univariate unit root processes. 

The Bayesian VAR with a Minnesota prior was shown by \cite{Robertson} to produce forecasts superior to the conventional VAR, univariate models, and traditional simultaneous equation models. However, this approach is very restrictive; in particular, it assumes that all series are contemporaneously uncorrelated, and it requires the specification of several hyperparameters.  Moreover, the Minnesota prior cannot accommodate large VARs itself.  As pointed out by \cite{BGR}, when constructing a 40 variable system, in addition to the Minnesota prior, \cite{Litterman1986} imposes strict economically-motivated restrictions to limit the number of variables in each equation.

Modern Bayesian VAR extensions originally proposed in \cite{kadiyala} and compiled by \cite{Koop} show that empirical regularization methods alone allow for the accurate forecasting of large VARs.  Such procedures impose data-driven restrictions on the parameter space while allowing for more general covariance specifications and estimation of hyperparameters via empirical Bayes or Markov chain Monte Carlo methods.  These approaches are computationally expensive, and multi-step forecasts are nonlinear and must be obtained by additional simulation.  Using a conjugate Gaussian-Wishart prior, \cite{BGR} extend the Minnesota prior to a high-dimensional setting with a closed-form posterior distribution.  Their technique uses a single hyperparameter, which is estimated by cross-validation.  However, their specification does not perform variable selection, and their penalty parameter selection procedure is more natural within a frequentist framework.

More recent attempts to reduce the parameter space of VARs have incorporated the \emph{lasso} \citep{tibs}, a least squares variable selection technique.  These approaches include the \emph{lasso-VAR} proposed by \cite{hsu} and further explored by \cite{BickelSong}, \cite{li}, and \cite{Davis}.  Theoretical properties were investigated by \cite{Kock} and by \cite{basu}.  \cite{Gefang} considers a Bayesian implementation of the elastic net, an extension of the lasso proposed by \cite{Hastie} that accounts for highly correlated covariates.  However, their implementation is not computationally tractable and they do not observe much of a forecasting improvement over existing methods.  The lasso-VAR has several advantages over Bayesian approaches as it is more computationally efficient in high dimensions, performs variable selection, and can readily compute multi-step forecasts and their associated prediction intervals.    

In many applications, a VAR's forecasts can be improved by incorporating unmodeled exogenous variables, which are determined outside of the VAR.  Examples of exogenous variables are context-dependent and range from leading indicators to weather-related measurements.  In many scenarios, global macroeconomic variables, such as world oil prices, may be considered exogenous.  Such models are most commonly referred to as ``VARX''  in the econometrics literature, but they are also known as ``transfer function'' or ``distributed lag'' models.  
  
VARX has become an especially popular approach in the modeling of small open economies, as they are generally sensitive to a wide variety of global macroeconomic variables which evolve independently of their internal indicators.  For example \cite{cushman} use a structural VARX model to analyze the effect of monetary policy shocks in Canada.  The VARX is also amenable under scenarios in which forecasts are desired only from a subset of the included series in a VAR, as by construction its corresponding VARX has a reduced parameterization.  VARX models have received considerable attention not just in economics, but also marketing \citep{retail}, political science \citep{wood}, and real estate \citep{brooks}.  

Unfortunately, dimensionality issues have limited the utility of the VARX.  As a result of the aforementioned overparameterization concerns, in the conventional unrestricted VAR context most applications are limited to no more than 6 series \citep{Bernanke}, forcing the practitioner to specify \emph{a priori} a reduced subset of series to include.
The VARX faces similar restrictions.
As outlined in \cite{Penm},
lag order, the maximum number of lagged observations to include, may differ between modeled and unmodeled series.  Hence, in order to select a VARX model using standard information-criterion minimization based methods, one must fit all subset models up to the predetermined maximal lag order for both the series of forecasting interest (which we refer to as \emph{endogenous} throughout this paper) and the exogenous series.  Moreover, unlike the conventional VAR, zero constraints (restrictions fixing certain model parameters to zero) are generally expected.

As it is often viewed as an economic rather than statistical problem, reducing the parameter space of the VARX model has received considerably less attention. \cite{ocampo} extend the aforementioned Bayesian VAR estimation methods to the VARX context.  \cite{george} apply stochastic search variable selection to the VARX framework; it provides a data-driven method to determine zero restrictions, but their approach is not scalable to high dimensions.  \cite{chiuso} propose estimating a VARX model with lasso and group lasso penalties but do not elaborate on potential group structures.  

This paper seeks to bridge the considerable gap between the regularization and macroeconomic forecasting communities.  We develop the VARX-L framework which allows for high-dimensional penalized VARX estimation while incorporating the unique structure of the VARX model in a computationally efficient manner.  In order to implement this framework, we develop substantial modifications to existing lasso and group lasso solution algorithms, which were designed primarily for univariate regression applications with no time dependence.
  
We extend the lasso and its structured counterparts to impose structured sparsity on the VARX, taking into account characteristics such as lag coefficient matrices, the delineation between a component's own lags and those of another component, and a potential nested structure between endogenous and exogenous variables.  Our methods offer great flexibility in capturing the underlying dynamics of an economic system while imposing minimal assumptions on the parameter space.  

Moreover, unlike previous approaches, due to our adaptation of convex optimization algorithms to a multivariate time series setting, our models are well-suited for the simultaneous forecasting of high-dimensional low-frequency macroeconomic time series.  In particular, our models allow for prediction under scenarios in which the number of component series and included exogenous variables is close to or exceeds the length of the series.  Our procedures, which avoid the use of subjective or complex hyperparameters, are publicly available in our {\tt R} package {\tt BigVAR} and can be easily applied by practitioners.

Section \ref{Sec2} describes the notation used throughout the paper and introduces our structured regularization methodology.

Section \ref{sec6} provides our implementation details and presents three macroeconomic forecasting applications. Section \ref{sec8} details the ``Minnesota VARX-L,'' an extension that allows for the incorporation of unit root nonstationarity by shrinking toward a vector random walk, section \ref{sec52} presents a simulation study, and section \ref{sec7} contains our conclusion.  The appendix details the solution strategies and algorithms that comprise the VARX-L class of models.

\section{Methodology}
\label{Sec2}

A $k$-dimensional multivariate time series $\{ \mathbf{y}_t \}_{t=1}^T$ and $m$-dimensional exogenous multivariate time series $\{\mathbf{x}_t\}_{t=1}^{T}$ follow a vector autoregression with exogenous variables of order $(p,s)$, denoted $\text{VARX}_{k,m}(p,s),$ if the following linear relationship holds (conditional on initialization): 
\begin{align}
\label{VAR1a}
\mathbf{y}_t=\utwi{\nu}+\sum_{\ell=1}^p\PhiB^{(\ell)}\mathbf{y}_{t-\ell}+\sum_{j=1}^s \betaB^{(j)}\mathbf{x}_{t-j}+\mathbf{u}_t \; \text{ for } \;t=1,\ldots,T,
\end{align}
in which $\utwi{\nu}$ denotes a $k$-dimensional constant intercept vector,  
$\PhiB^{(\ell)}$ represents a $k\times k$ endogenous coefficient matrix at lag $\ell=1,\dots,p$, $\betaB^{(j)}$ represents a $k\times m$ exogenous coefficient matrix at lag $j=1,\dots,s$,
and 
$\mathbf{u}_t$ denotes a $k$-dimensional white noise vector that is independent and identically distributed with mean zero and nonsingular covariance matrix $\mathbf{\Sigma}_u$.  A VAR, which is a special case of the VARX, can be represented by Equation \eqref{VAR1a} with $\betaB^{(j)}=\utwi{0}$ for $j=1,\dots,s$.

In a low-dimensional setting, in which the number of included predictors is substantially smaller than the length of the series, $T$, the VARX model can be fit 
by multivariate least squares, with $\utwi{\nu},\PhiB=[\PhiB^{(1)},\dots,\PhiB^{(p)}]$, and $\betaB=[\betaB^{(1)},\dots,\betaB^{(s)}]$ estimated as
\begin{align}
\argmin_{\utwi{\nu},\PhiB,\betaB} \quad  \sum_{t=1}^T\|\mathbf{y}_t-\utwi{\nu}-\sum_{\ell=1}^p\PhiB^{(\ell)}\mathbf{y}_{t-\ell}-\sum_{j=1}^s \betaB^{(j)}\mathbf{x}_{t-j}\|_F^2\label{eq:ls},
\end{align}
in which $\|A\|_F=\sqrt{\sum_{i,j}A_{ij}^2}$ denotes the Frobenius norm of a matrix $A$, which reduces to the $L_2$ norm when $A$ is a vector. 
In the absence of regularization, the $\text{VARX}_{k,m}(p,s)$ requires the estimation of $k(1+kp+ms)$ regression parameters.  In the following section, we will apply several convex penalties to Equation \eqref{eq:ls} which aid in reducing the parameter space of the VARX by imposing sparsity in $\PhiB$ and $\betaB$.

\subsection{VARX-L: Structured Penalties for VARX Modeling}
\label{sec3}
In this section, we introduce VARX-L, a general penalized multivariate regression framework for large VARX models.  We consider structured objectives of the form
\begin{align}
\label{PenFunForm}
\min_{\utwi{\nu},\PhiB,\betaB}  \sum_{t=1}^T\|\mathbf{y}_t-\utwi{\nu}-\sum_{\ell=1}^p\PhiB^{(\ell)}\mathbf{y}_{t-\ell}-\sum_{j=1}^s \betaB^{(j)}\mathbf{x}_{t-j}\|_F^2+\lambda \bigg(\mathcal{P}_y(\PhiB)+ \mathcal{P}_x(\betaB)\bigg),
\end{align}
in which $\lambda\geq 0$ is a penalty parameter, which is selected in a sequential, \emph{rolling} manner according to a procedure that is discussed in section \ref{sec4}, $\mathcal{P}_y(\PhiB)$ denotes a penalty function on endogenous coefficients,
 and $\mathcal{P}_x(\betaB)$  denotes a penalty function on exogenous coefficients.

Table \ref{tab:tabGS} details the penalty structures proposed in this paper; all but the last have this separable structure.  In the following section, we will discuss each penalty structure in detail.  Note that since we utilize a single penalty parameter for all model coefficients, it is required that all included series are on the same scale; hence we assume that prior to estimation, the series are standardized to each have zero mean and unit variance.  

     \begin{table}
\fontsize{6}{7.2}\selectfont
    \caption{ \fontsize{6}{7.2}\selectfont
 \label{tab:tabGS}  The Proposed VARX-L Penalty Functions.  Note that $\PhiB_{\text{on}}^{(\ell)}$ and $\PhiB_{\text{off}}^{(\ell)}$ denote the diagonal and off-diagonal elements of coefficient matrix $\PhiB^{(\ell)}$, respectively.
    }  
   \begin{tabular}{ l | c| c }
Group Name& $\mathcal{P}_y(\PhiB)$ & $\mathcal{P}_x(\betaB)$ \\
    \hline
\refstepcounter{equation}(\theequation) \label{VARX1} Lag &  $\sqrt{k^2}\sum_{\ell=1}^p\|\PhiB^{(\ell)}\|_F$ &  $\sqrt{k}\sum_{j=1}^s\sum_{i=1}^m\|\betaB_{\cdot,i}^{(j)}\|_F$  \\
\refstepcounter{equation}(\theequation) \label{VARX2} Own/Other& $\sqrt{k}\sum_{\ell=1}^p ||\PhiB_{\text{on}}^{(\ell)}||_F+\sqrt{k(k-1)}\sum_{\ell=1}^p||\PhiB_{\text{off}}^{(\ell)}||_F$ &$\sqrt{k}\sum_{j=1}^s\sum_{i=1}^m\|\betaB_{\cdot,i}^{(j)}\|_F$ \\
\refstepcounter{equation}(\theequation) \label{VARX3} Sparse Lag& $(1-\alpha)\sqrt{k^2}\sum_{\ell=1}^p\|\PhiB^{(\ell)}\|_F+\alpha\|\PhiB\|_1$ & $(1-\alpha)\sqrt{k}\sum_{j=1}^s\sum_{i=1}^m\|\betaB_{\cdot,i}^{(j)}\|_F$ +$\alpha\|\betaB\|_1$  \\
\refstepcounter{equation}(\theequation) \label{VARX4} Sparse Own/Other& $(1-\alpha)\big(\sqrt{k}\sum_{\ell=1}^p||\PhiB_{\text{on}}^{(\ell)}||_F+\sqrt{k(k-1)}\sum_{\ell=1}^p||\PhiB_{\text{off}}^{(\ell)}||_F\big)+\alpha\|\PhiB\|_1$ &  $(1-\alpha)\sqrt{k}\sum_{j=1}^s\sum_{i=1}^m\|\betaB_{\cdot,i}^{(j)}\|_F$ +$\alpha\|\betaB\|_1$\\
\refstepcounter{equation}(\theequation) \label{VARX5} Basic & $\|\PhiB\|_1$  & $\|\betaB\|_1$ \\  
\hline\\
\refstepcounter{equation}(\theequation) \label{VARX6} Endogenous-First & \multicolumn{2}{c}{$\mathcal{P}_{y,x}(\PhiB,\betaB)=\sum_{\ell=1}^p\sum_{j=1}^k\bigg(\|[\PhiB_{j,\cdot}^{(\ell)},\betaB_{j,\cdot}^{(\ell)}]\|_F+\|\betaB_{j,\cdot}^{(\ell)}\|_F\bigg)$}   
  \end{tabular}
 \end{table}

Equations \eqref{VARX1}-\eqref{VARX2} adapt the \emph{group lasso} penalty proposed by \cite{yuan} to the VARX setting.  The group lasso partitions the parameter space into groups of related variables which are shrunk toward zero.  Within a group, all variables are either identically set to zero or are all nonzero.  Our choices of $\mathcal{P}_y$ and $\mathcal{P}_x$ create structured sparsity based on pre-specified groupings, which are designed to incorporate the intrinsic lagged structure of the VARX.  
The proposed ``lag group'' methods have a substantial advantage over popular Bayesian approaches in that they will both shrink least squares estimates toward zero as well as perform variable selection in a computationally efficient manner.

Sparsity in the coefficient matrix is desirable when $k$ and $m$ are large because the conventional VARX is grossly overparameterized.  As stated in \cite{Litterman1984}, it is widely believed in macroeconomic forecasting that small bits of relevant information exist throughout the data, and economic theory is not informative with regard to the manner in which this information is scattered.  The proposed VARX-L framework provides a systematic approach to filter as much information as possible, assigning each bit an appropriate weight.

The group lasso penalty function was explored in the VAR context by \cite{BickelSong} who consider several structured groupings with a particular emphasis on creating a distinction between a series' own lags and those of another series.  Theoretical properties of the use of a group lasso penalty in the VAR setting were further explored by \cite{mich1}.  

A feature of the Lag Group VARX-L is that it does not impose sparsity within a group.  \cite{BickelSong} attempt to circumvent this constraint by including several additional lasso penalties, but such an approach requires a multi-dimensional gridsearch to select penalty parameters.  The penalties for the proposed Sparse Group VARX-L and Sparse Own/Other Group VARX-L, listed in Equations \eqref{VARX3}-\eqref{VARX4}, instead implement the \emph{sparse group lasso} \citep{simon}, which extends the group lasso to allow within-group sparsity.  The sparse group lasso can be viewed analogously to the elastic net \citep{Hastie} extended to structured penalties. 

The penalty for the Basic VARX-L adapts the lasso \eqref{VARX5}; it considers no structure, or can be viewed as a group lasso penalty that assigns each coefficient to a singleton group.  In very high-dimensional scenarios, this most basic penalty has computational advantages as compared to more complex approaches.  Finally, the penalty for the proposed Endogenous-First VARX-L, Equation \eqref{VARX6}, incorporates a nested penalty structure such that, within a lag, endogenous coefficients are prioritized before their exogenous counterparts.  Since this penalty structure is not separable in the manner of Equation \eqref{PenFunForm}, its penalty function is denoted as $\mathcal{P}_{y,x}$.

\subsubsection*{Group VARX-L }
\begin{figure}
\centering
\caption{ \label{fig:figGL} \footnotesize Example sparsity pattern (active elements shaded) produced by a {\bf Lag Group} $\text{VARX-L}_{3,2}(5,3)$}
\includegraphics[scale=.7]{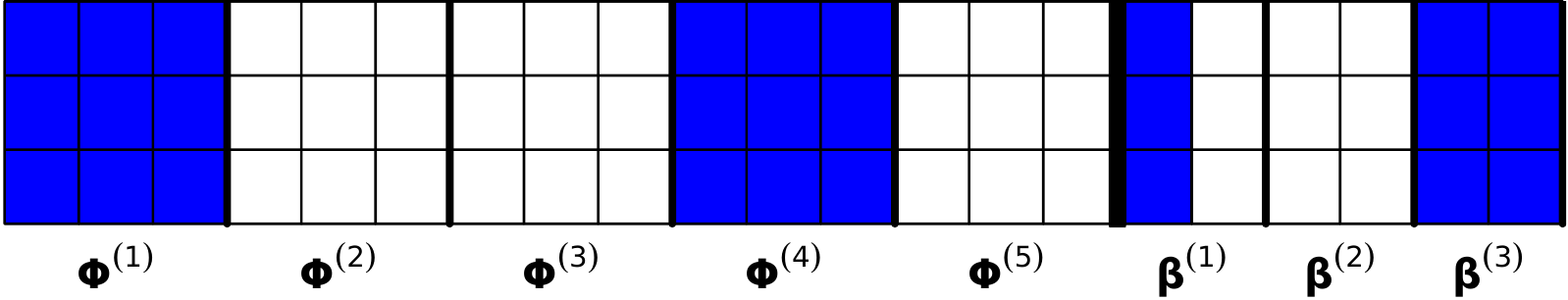}
\end{figure}

We first present the \emph{Lag} Group VARX-L \eqref{VARX1}, in which the endogenous coefficients are grouped according to their lagged coefficient matrix $\PhiB^{(\ell)}$ for $\ell=1,\dots,p$, and at every lag, each exogenous component series is partitioned into its own group.  This structured grouping is expressed as  
\begin{align*}
\mathcal{P}_{y}(\PhiB)= \sqrt{k^2}\sum_{\ell=1}^p\|\PhiB^{(\ell)}\|_F,\quad \mathcal{P}_{x}(\betaB)=\sqrt{k}\sum_{j=1}^s\sum_{i=1}^m\|\betaB_{\cdot,i}^{(j)}\|_F.  
\end{align*}
Note that since the endogenous and exogenous groups differ in cardinality, it is required to weight the penalty to avoid regularization favoring larger groups.  This structure implies that for each endogenous series, a coefficient matrix at lag $\ell$ is either entirely nonzero or entirely zero.
Similarly, the relationship between an exogenous and endogenous series at lag $j$ will either be nonzero for all endogenous series or identically zero.  A potential sparsity pattern generated by this structure (with $k=3$, $p=5$, $m=2$, and $s=3$) is shown in Figure \ref{fig:figGL} with the active (i.e. nonzero) elements shaded.  

In comparison to Bayesian regularization methods, such as stochastic search variable selection \citep{george}, estimating the Lag Group VARX-L is tractable even in high dimensions.
We are able to extend the efficient group lasso solution method proposed by \cite{goldfarb}, who utilize a block coordinate descent procedure and transform each ``one group'' subproblem to a trust-region framework.  These subproblems can then be solved efficiently via univariate optimization.  Details of our algorithm are provided in section \ref{grouplassosol} of the appendix.

The Lag Group structure is advantageous for applications in which all series tend to exhibit comparable dynamics, such as forecasting the disaggregate subcomponents of a composite index.  It also can serve as a powerful tool for lag selection.  However, in many settings, it may not be appropriate to give equal consideration to every entry in a coefficient matrix.  Diagonal entries of each $\PhiB^{(\ell)}$, which represent regression on a series' own lags, are in many applications more likely to be nonzero than are off-diagonal entries, which represent lagged cross-dependence with other components.  The \emph{Own/Other} Group VARX-L (\ref{VARX2}) allows for the partitioning of each lag matrix $\PhiB^{(\ell)}$ into separate groups by assigning the endogenous penalty structure
\begin{align*}
\mathcal{P}_y(\PhiB)=\sqrt{k}\sum_{\ell=1}^p ||\PhiB_{\text{on}}^{(\ell)}||_F+\sqrt{k(k-1)}\sum_{\ell=1}^p||\PhiB_{\text{off}}^{(\ell)}||_F, 
\end{align*}
in which $\PhiB_{\text{on}}^{(\ell)}$ denotes the diagonal elements of $\PhiB^{(\ell)}$ and $\PhiB_{\text{off}}^{(\ell)}$ denotes its off-diagonal entries.

An example of this sparsity pattern is shown in Figure \ref{fig:figGLOO}.  The computational modifications required to utilize the Own/Other structure are detailed in section \ref{grouplassooosol} of the appendix.  This delineation between own lags and other lags is often incorporated in macroeconomic forecasting.  As detailed in \cite{Litterman5yr}, the traditional Minnesota prior operates under the assumption that a series' own past values account for most of its variation, hence they are shrunk by a smaller factor than realizations of other series.  The strong forecasting performance of the VARX-L procedures that utilize the Own/Other structure in section \ref{sec53} provides further justification for Litterman's beliefs.

In addition to the Own/Other and Lag group structures, one could consider a variety of application-dependent partitions.  This idea was briefly explored in \cite{BickelSong} who posit potentially segmenting financial or economic series based upon industry or sector characteristics in addition to a lag based grouping.  However, since in many applications we lack \emph{a priori } information about the potential relationships of the included series, we prefer to structure our groupings around the intrinsic structure of the VARX.

\begin{figure}
\centering
\label{fig:figGLOO}
\caption{ \label{fig:figGLOO}\footnotesize{Example sparsity pattern (active elements shaded) produced by an {\bf Own/Other Group} $\text{VARX-L}_{3,2}(5,3)$}}
\includegraphics[scale=.7]{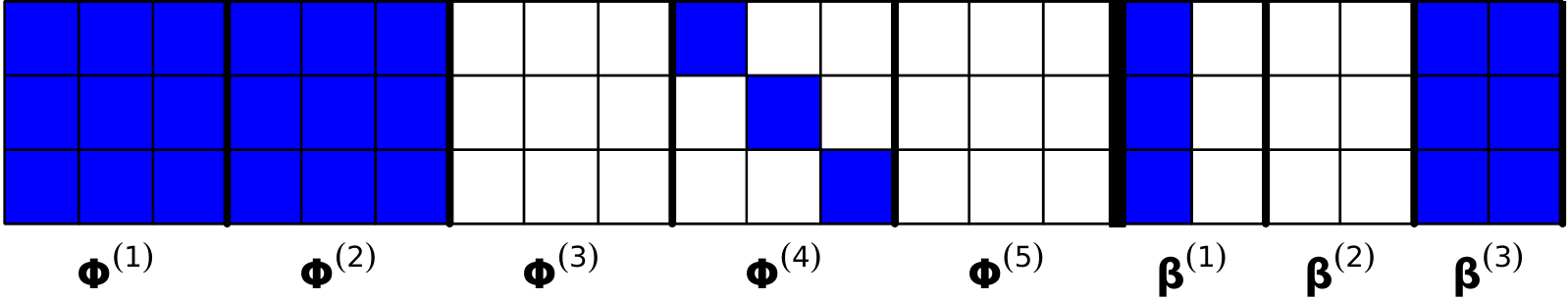}  
\end{figure}

\subsubsection*{Sparse Group VARX-L}
For certain applications, a group penalty might be too restrictive. If a group is active, all coefficients in the group will be nonzero, and including a large number of groups substantially increases computation time. Moreover, it is inefficient to include an entire group if, for example, only one coefficient is truly nonzero.  The \emph{sparse group lasso}, proposed by \cite{simon} allows for within-group sparsity through a convex combination of lasso and group lasso penalties.  The Sparse Lag Group VARX-L \eqref{VARX3} results in a penalty of the form
\begin{align*}
&\mathcal{P}_y(\PhiB)=(1-\alpha)\big(\sqrt{k^2}\sum_{\ell=1}^p\|\PhiB^{(\ell)}\|_F\big)+\alpha\|\PhiB\|_{1},\quad \mathcal{P}_x(\betaB)=(1-\alpha)\big(\sqrt{k} \sum_{j=1}^{s}\sum_{i=1}^m\|\betaB_{\cdot,i}^{(j)}\|_F\big)+\alpha\|\betaB\|_{1},
\end{align*}
in which $0\leq \alpha\leq 1$ is an additional penalty parameter that controls within-group sparsity.  Without prior knowledge of the important predictors, we weight according to relative group sizes and set $\alpha=\frac{1}{k+1}$, though $\alpha$ could also be estimated by cross-validation.

The inclusion of the $L_1$ norm allows for within-group sparsity, hence even if a group is considered active individual coefficients within it can be set to zero.  An example sparsity pattern is depicted in Figure \ref{fig:figSGL}.

\begin{figure}
\centering
\label{fig:figSGL}
\caption{ \label{fig:figSGL} \footnotesize Example sparsity pattern (active elements shaded) produced by a {\bf Sparse Lag Group} $\text{VARX-L}_{3,2}(5,3)$}
\includegraphics[scale=.7]{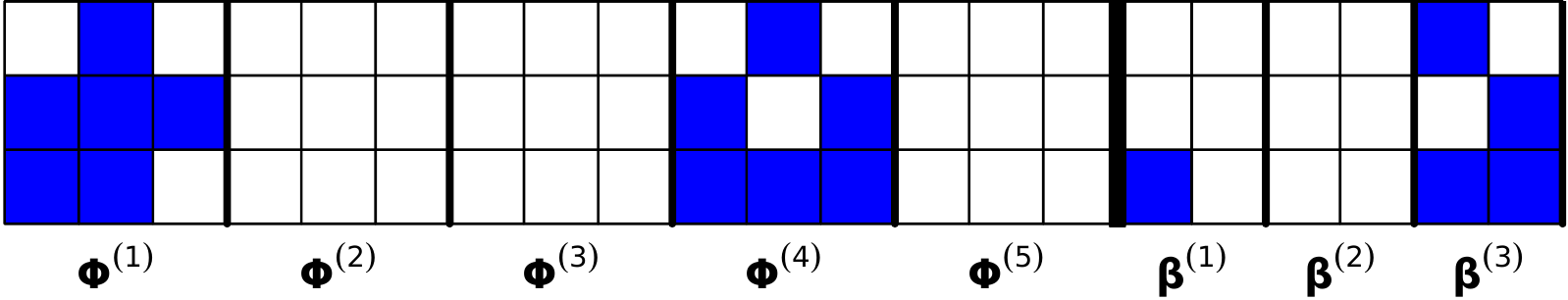}. 
\end{figure}

Since the inclusion of within-group sparsity does not create a separable objective function, conventional group lasso solution methods, such as coordinate descent, are no longer applicable.  Following \cite{simon}, our estimation algorithm for the Sparse Lag Group VARX-L makes use of proximal gradient descent.
The details of this approach and our implementation are provided in section \ref{sglsol} of the appendix.  This penalty can be extended to alternative groupings.  Consequently, we also consider the \emph{Sparse Own/Other Group} VARX-L \eqref{VARX4} as an estimation procedure.

\subsubsection*{Basic VARX-L}
\begin{figure}
\centering
\caption{ \footnotesize \label{fig:fig1} Example sparsity pattern (active elements shaded) produced by a {\bf Basic} $\text{VARX-L}_{3,2}(5,3)$}
\includegraphics[scale=.7]{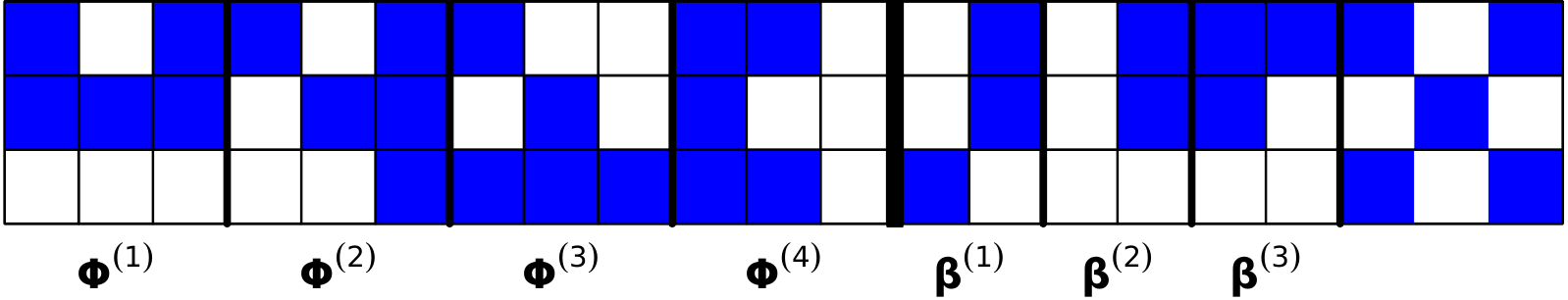}  
  \end{figure}

The Basic VARX-L \eqref{VARX5}, proposed by \cite{chiuso}, incorporates no structure and can be viewed as a special case of the Sparse Lag Group VARX-L in which $\alpha=1$, resulting in penalties of the form
\begin{align*}
&\mathcal{P}_y(\PhiB)=\|\PhiB\|_1,\quad \mathcal{P}_x(\betaB)=\|\betaB\|_1.  
\end{align*}
The $L_1$ penalty will induce sparsity in the coefficient matrices $\PhiB$ and $\betaB$ by zeroing individual entries.  An example sparsity pattern is depicted in Figure \ref{fig:fig1}.

A major advantage of the Basic VARX-L over its structured counterparts is its computational tractability.
Our solution approach
involves the use of coordinate descent, popularized for lasso problems by \cite{friedman}.  Coordinate descent consists of partitioning the Basic VARX-L into subproblems for each scalar element $[\PhiB,\betaB]_{ij}$, solving component-wise, then updating until convergence.  This approach is computationally efficient since, in the Basic VARX-L context, each subproblem has a closed-form solution.  \cite{tseng} establishes that global convergence arises from solving individual subproblems in the coordinate descent framework.  Our solution strategy and algorithm are detailed section \ref{lassovarsol} of the appendix.

\subsection{An Endogenous-First Active Set}\label{sec3.3}
We have previously only considered structures that assign endogenous and exogenous variables to separate groups.  In this section, we consider a nested structure that can take into account the relative importance between endogenous and exogenous predictor series.
 
In certain scenarios, there may exist an \emph{a priori} importance ranking among endogenous and exogenous variables.  For example, the endogenous variables could represent economic indicators of interest in a small open economy, with global macroeconomic indicators serving as exogenous variables.  In such a scenario, it may be desirable for exogenous variables to enter into a forecasting equation only if endogenous variables are also present at a given lag $\ell$.  We can consider such a structure by utilizing a \emph{hierarchical group lasso} penalty (see, e.g. \cite{Jenatton}).  The Endogenous-First VARX-L penalty function \eqref{VARX6} takes the form
\begin{align}
\label{EF1}
\mathcal{P}_{y,x}(\PhiB,\betaB)=\sum_{\ell=1}^p\sum_{j=1}^k\big(\|[\PhiB_{j,\cdot}^{(\ell)},\betaB_{j,\cdot}^{(\ell)}]\|_F+\|\betaB_{j,\cdot}^{(\ell)}\|_F\big).
\end{align}
Under this structure, at a given lag, exogenous variables can enter the model only after the endogenous variables at the same lag.  Note that this structure requires that $s\leq p$.  It should also be noted that \eqref{EF1} decouples across rows, allowing for separate nested structures across each endogenous series.  This sparsity pattern is depicted in Figure \ref{fig:hvar12}.

\begin{figure}
\centering
\label{fig:hvar12}
\includegraphics[scale=.5]{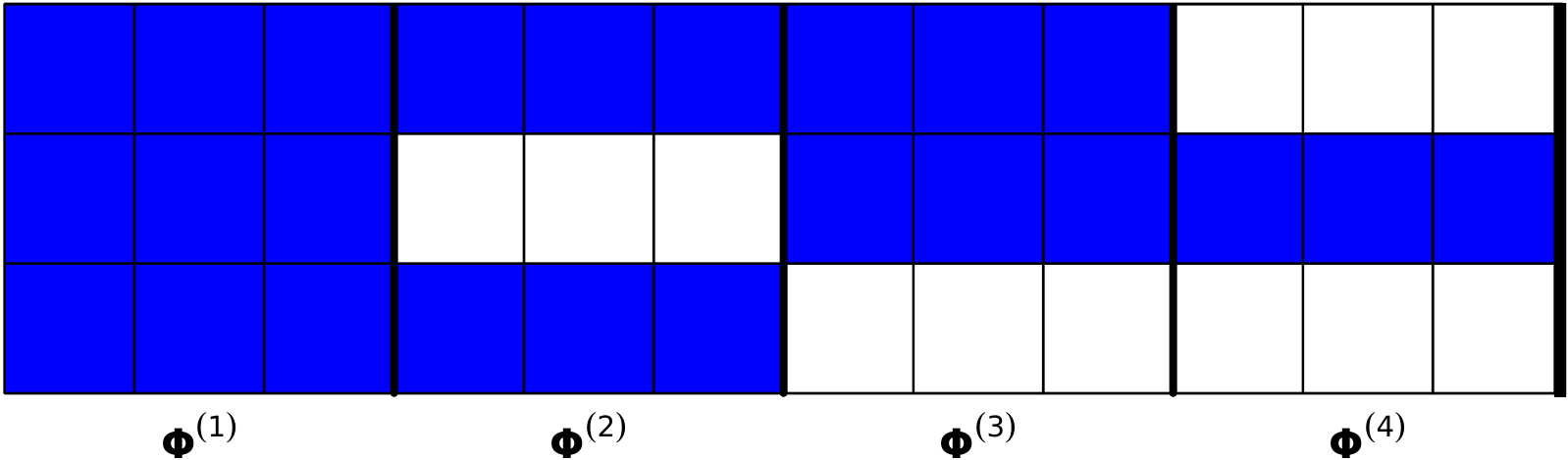}
\includegraphics[scale=.5]{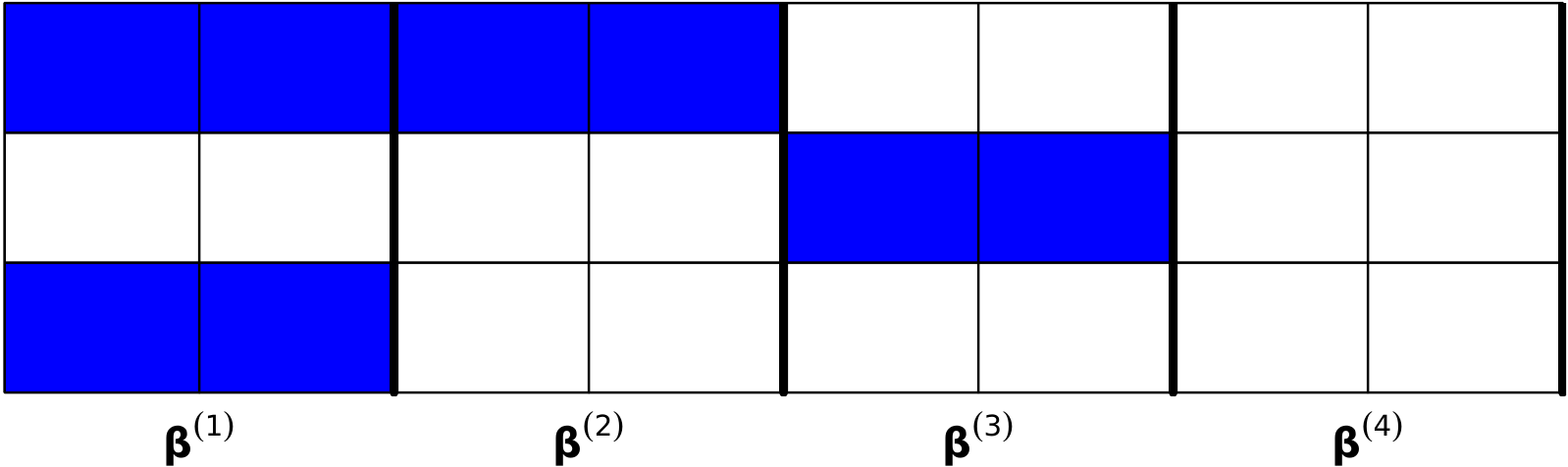}
\caption{ \footnotesize \label{fig:hvar12} Example sparsity pattern (active elements shaded) generated by an {\bf Endogenous-First} $\text{VARX-L}_{3,2}(4,4)$.  Note that a row in $\betaB^{(\ell)}$ can only be nonzero if the corresponding row in $\PhiB^{(\ell)}$ is also nonzero. }
\end{figure}

Most group lasso solution methods, such as block coordinate descent, take advantage of the separability of groups to improve computational performance.  Although the nested structure is not directly separable, based on the methodology of \cite{Jenatton}, its dual can be solved in one pass of block coordinate descent.  Details of the solution approach and our algorithm are provided in section \ref{HVARX1} of the appendix.

\section{High-Dimensional Macroeconometrics}
\label{sec6}

In this section, we evaluate our regularization procedures in three macroeconomic data applications: two high-dimensional and one low-dimensional.  In our first two applications, we consider applying the proposed VARX-L procedures on the widely used set of US macroeconomic indicators originally constructed by \cite{stockdataset}.  Our second example considers forecasting a small set of Canadian macroeconomic indicators and incorporating the previous US data as exogenous series.  Section \ref{sec4} outlines the practical implementation of our penalty parameter selection procedure, section \ref{sec5.1} describes the benchmarks that we compare our models against, section \ref{sec53} details our macroeconomic applications, and section \ref{secMCS} provides an expanded comparison of the relative forecasting performance among the competing models using the model confidence sets framework developed by \cite{Hansen2012}.

\subsection{Practical Implementation}
\label{sec4}
The regularization parameter, $\lambda$, is not known in practice and is typically chosen via cross-validation.  In this section, we detail our strategy for selecting $\lambda$.  Following \cite{friedman}, we select from a grid of potential penalty parameters that starts with the smallest value in which all components of $[\PhiB,\betaB]$ will be zero and decreases in log-linear increments.  
This value differs for each procedure and can be inferred by their respective algorithms.  The starting values are summarized in Table \ref{tab:tabSP} located in section \ref{pengrid} of the appendix.  The number of gridpoints, $N$, as well as the depth of the grid are left to user input.  A deep grid and large number of gridpoints result in increased computational costs and often do not improve forecasting performance.  We have found that a grid depth $\frac{1}{25}\lambda_{\max}$ and 10 gridpoints achieve adequate forecast performance in most scenarios.

Due to time-dependence, our problem is not well-suited to traditional $K$-fold cross-validation.  Instead, following \cite{BGR}, we propose choosing the optimal penalty parameter by minimizing $h$-step ahead mean-square forecast error (MSFE), in which $h=1,2,3,\dots$ denotes the desired forecast horizon.  We divide the data into three periods: one for initialization, one for training, and one for forecast evaluation.  Define time indices $T_1=\left \lfloor \frac{T}{3} \right\rfloor,  T_2=\left\lfloor \frac{2T}{3} \right\rfloor$.

We start our validation process by fitting a model using all data up to time $T_1$ and forecast $\hat{\mathbf{y}}_{T_1+1}^{\lambda_i}$ for $i=1,\dots,N$.  We then sequentially add one observation at a time and repeat this process until time $T_2-h$.  This procedure is illustrated in Figure \ref{fig:figcv}.

\begin{figure}[H]
\centering
\label{fig:figcv}
\includegraphics[scale=1]{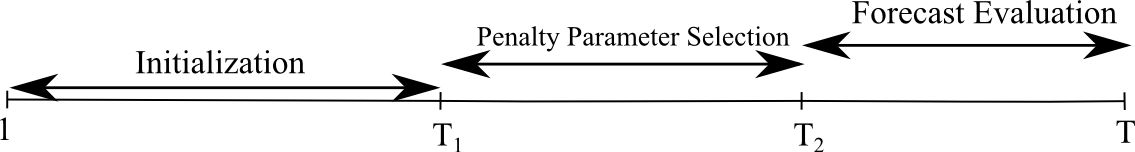}
\caption{ \footnotesize \label{fig:figcv} Illustration of Rolling Cross-Validation}
\end{figure}

We select $\hat{\lambda}$ as the minimizer of
\begin{align}
\label{PEN1}
  MSFE(\lambda_i)=\frac{1}{(T_2-T_1-h+1)}\sum_{t=T_1}^{T_2-h} \|\hat{\mathbf{y}}_{t+h}^{\lambda_i}-\mathbf{y}_{t+h}\|_F^2.
\end{align}
Finally, we evaluate the $h$-step ahead forecast accuracy of $\hat{\lambda}$ from time origins $t=T_2-h$ to $t=T-h$.  If desired, additional criterion functions can be substituted.  MSFE is the most natural criterion given our use of the least squares objective function.  Rather than parallelizing the cross-validation procedure, our approach uses the result from the previous period as an initialization or ``warm start,'' which substantially decreases computation time.  The penalty parameter selection procedure is presented in Algorithm \ref{alg2} in the appendix.

\subsubsection*{Multi-step Predictions}

The VARX-L framework can easily accommodate multi-step ahead forecasting.  To do so,
we modify our solution algorithms to calculate \emph{direct} multi-step ahead forecasts.  Essentially, computing direct $h$-step forecasts involves solving the standard VARX-L objective \eqref{PenFunForm} while leaving a gap of $h$ observations.  
\begin{align*}
\min_{\utwi{\nu},\PhiB,\betaB}  \sum_{t=h}^{T}\|\mathbf{y}_t-\utwi{\nu}-\sum_{\ell=1}^p\PhiB^{(\ell)}\mathbf{y}_{t-h-\ell+1}-\sum_{j=1}^s \betaB^{(j)}\mathbf{x}_{t-h-j+1}\|_F^2+\lambda \bigg(\mathcal{P}_y(\PhiB)+ \mathcal{P}_x(\betaB)\bigg).\\
\end{align*}
Per \cite{clark}, the direct $h$-step ahead forecast can be calculated as
\begin{align*}
 \hat{\mathbf{y}}_{t+h}=\hat{\utwi{\nu}}+\widehat{\PhiB}^{(1)}\mathbf{y}_{t}+\dots+\widehat{\PhiB}^{(p)}\mathbf{y}_{t-p+1}+\widehat{\betaB}^{(1)}\mathbf{x}_t+\dots+\widehat{\betaB}^{(s)}\mathbf{x}_{t-s+1}. 
\end{align*}

The \emph{iterative} approach, in which multi-step forecasts are computed recursively as $1-$step ahead forecasts using predicted values is another popular technique to compute long-horizon forecasts.  This approach is not directly extendable to the VARX setting, as we do not forecast the exogenous series.

If iterative multi-step predictions are desired, one could instead fit the full $\text{VAR}_{k+m}$.  However, as shown in \cite{swforecast}, direct forecasts are more robust to model misspecification, making them more appropriate in high-dimensional settings.  As our macroeconomic applications consider quarterly data, in section \ref{sec53} we compute both $1-$step and $4-$step ahead forecasts.

\subsubsection*{Selecting a Structure}

The VARX-L framework offers many choices for structured penalties suited toward a wide range of applications.  Under scenarios in which little is known about the potential dynamic dependence of the included series, the Basic VARX-L makes no underlying structural assumptions.

The Lag Group and Sparse Lag Group VARX-L structures are most appropriate when the endogenous series are closely related; for example, if the series of interest comprise unemployment rates segmented by state or census region.  The Own/Other Group and Sparse Own/Other Group VARX-L structures are most appropriate for macroeconomic applications in which a series' own lags are thought to have substantially different temporal dependence than those of ``other'' series.  As an example, one could consider a disparate group of series traditionally examined in small-scale macroeconometric forecasting applications: the US Federal Funds Rate, the GDP Growth Rate, and the Consumer Price Index.  The Endogenous-First structure is best suited toward applications in which the forecasting effectiveness of the exogenous series is unknown.

One potential diagnostic tool involves fitting the Sparse Lag Group VARX-L with both $\lambda$ and $\alpha$ selected according to rolling cross validation.  A selected value of $\alpha$ close to 0 indicates evidence of strong groupwise sparsity, while a value close to 1 indicates unstructured sparsity, and values in the middle provide evidence for some combination of the two.  

Since the computational time required to apply all procedures is manageable, in practice, we suggest fitting several VARX-L structures and selecting the approach that achieves the best out of sample forecasting performance.

\subsection{Methods for Comparison}
\label{sec5.1}
A conventional VARX model selection approach in a low-dimensional setting involves fitting a $\text{VARX}_{k,m}(\ell,j)$ by least squares for $0\leq \ell \leq p,$ $0\leq j \leq s$ and selecting lag orders for both the endogenous and exogenous series based on an information criterion, such as Akaike's Information Criterion (AIC) or Bayesian Information Criterion (BIC).  Per \cite{Lutk}, the AIC and BIC of a $\text{VARX}_{k,m}(\ell,j)$ are defined as
\begin{align*}
&\text{AIC}(\ell,j)=\log|\widehat{\Sigma}_{u}^{\ell,j}|+\frac{2\big(k(k\ell+mj)\big)}{T},\\  
&\text{BIC}(\ell,j)=  \log|\widehat{\Sigma}_{u}^{\ell,j}|+\frac{\log(T)\big(k(k\ell+mj)\big)}{T}.
\end{align*}
in which $\widehat{\Sigma}_{u}^{\ell,j}$ is the residual sample covariance matrix obtained from the estimated $\text{VARX}_{k,m}(\ell,j)$, and $|\Sigma|$ represents the determinant of $\Sigma$.  The selected lag orders $(\ell,j)$ are then chosen as the minimizer of AIC or BIC.  AIC penalizes each model coefficient uniformly by a factor of two whereas BIC scales penalties relative to series length.  Hence, when $T$ is large, BIC will tend to select more parsimonious models than AIC.

We compare our methods against least squares model selection procedures that utilize AIC and BIC to select lag orders.  Since we are considering high-dimensional applications, in which $\widehat{\Sigma}_u$ could be ill-conditioned, we construct our least squares estimates using a variation of the approach developed by \cite{neumaier}.  This procedure constructs the least squares estimates using a QR decomposition, which obviates the need for explicit matrix inversion.  In addition, following a heuristic proposed by \cite{Hansen}, we impose a
ridge penalty: $\big((k\cdot \ell+m\cdot j)^2+(k\cdot \ell+m\cdot j)+1\big)\epsilon_{\text{machine}}$ scaled by the column norms of the lagged series $\mathbf{y}_{t-1},\dots,\mathbf{y}_{t-p},\mathbf{x}_{t-1},\dots,\mathbf{x}_{t-s}$, in which $\epsilon_{\text{machine}}$ denotes machine precision.  This penalty ensures that the determinant of $\widehat{\Sigma}_u$ is well-defined without noticeably impacting degree of freedom calculations.  However, in very high-dimensional settings, where $k\big(kp+ms) > kT$, the covariance $\Sigma_u$ is not well defined, hence information criterion based methods are not appropriate.  The specifics of our information criterion minimization routine are detailed in \cite{BigVAR}.    
   
We additionally compare our methods against two naive approaches that provide insight with regard to the level of temporal dependence in the data.  We first consider the unconditional {\em sample mean}, which will make $h$-step ahead forecasts at time $t+h$ based upon the average of all observed data up to time $t$:
$\hat{\bf y}_{t+h}=\frac1{t}\sum_{i=1}^{t}{\bf y}_{i}$.  Strong relative performance of the sample mean indicates that the sophisticated models of interest have low predictive power relative to a white noise process; the cause of which could be a number of factors, including weak temporal dependence or dependence of a nonlinear nature.
 
Second, we consider the vector {\em random walk} model, which makes $h$-step ahead forecasts based upon the most recent realization of the series, i.e.
$\hat{{\bf y}}_{t+h}={\bf y}_{t}$.  Superior performance of the vector random walk indicates high persistence or a strong degree of temporal dependence, as is often observed in macroeconomic data. 

Finally, we compare against two more sophisticated approaches.  First, we consider the popular Bayesian VAR with a modified Minnesota Prior proposed by \cite{BGR} (henceforth BGR).  Their approach acts very similarly to ridge regression in that it shrinks least squares coefficients toward zero with the degree of regularization determined by a single penalty parameter.  This parameter is chosen according to rolling cross validation as described in section \ref{sec4}.  As in \cite{BGR}, instead of fitting a $\text{VARX}_{k,m}(p,s)$, we fit a $\text{VAR}_{k+m}(p)$ and select the regularization parameter as the minimizer of $h-$step ahead MSFE across the $k$ endogenous series.  This allows BGR's method to make forecasts utilizing information from both the endogenous and exogenous series, and creates a direct comparison with our VARX-L framework.   

BGR's approach modifies the Minnesota Prior to make it computationally tractable in high dimensions, but it does not does not return sparse solutions.  Superior performance of the VARX-L methods relative to BGR's approach provides evidence as to the importance of imposing sparsity in obtaining accurate forecasts.  Details of our implementation of BGR's procedure are provided in section \ref{BGRAPP} in the appendix.  

In addition, we compare against a factor model \citep{anderson,stockbc}.  In a manner similar to principal components analysis, a factor model attempts to find a low-rank structure in the data that adequately accounts for a high percentage of the variation across the series.  Our implementation forecasts using a reduced-rank structure that accounts for at least 95 percent of the total variance of the included series.  For implementation details, consult \cite{ensor}.  We expect that the factor model will perform well under scenarios in which a small subset  of common factors drive the underlying dynamics of the larger cross-section of modeled series.

\subsection{ Macroeconometric Applications}
\label{sec53}
We evaluate our methods on the large and widely utilized macroeconomic dataset created by \cite{stockdataset} and later amended by \cite{Koop}.  The dataset consists of 168 quarterly US macroeconomic indicators containing information about various aspects of the economy, including income, industrial production, employment, stock prices, interest rates, exchange rates, etc.  The data ranges from Quarter 2 of 1959 to Quarter 3 of 2007 ($T=195$).  Per \cite{Koop}, the series can be categorized into several levels; we consider the following four:
 \begin{itemize}
   \item \emph{Small} ($k=3$): Three variables (Federal Funds Rate, Consumer Price Index, Gross Domestic Product growth rate). Core group, typically used in simple dynamic stochastic generalized equilibrium models;
\item \emph{Medium} ($k=20$): Small plus 17 additional variables containing aggregated economic information (e.g., consumption, labor, housing, exchange rates);
\item \emph{Medium-Large} ($k=40$): Medium plus 20 additional aggregate variables;
\item ``Large:'' Medium-Large plus 128 additional variables, consisting primarily of the components that make up the aggregated information (k=168). 
\end{itemize}

For a detailed description of each set of variables, consult \cite{Koop}.  As \cite{BGR} found that the greatest improvements in forecast performance occurred with the \emph{medium} VAR, that will be our initial focus.
We will attempt to forecast the \emph{medium} set of indicators ($k=20$) while using the additional variables from the \emph{medium-large} category as exogenous predictors ($m=20$).

Before estimation, each series is transformed to stationarity according to the specifications provided by \cite{stockdataset} and standardized by subtracting the sample mean and dividing by the sample standard deviation.
Quarter 2 of 1976 to Quarter 3 of 1992 is used for penalty parameter selection while Quarter 4 of 1992 to Quarter 3 of 2007 is used for forecast evaluation.  The MSFE (relative to the sample mean) over the evaluation period for each model is reported in Table \ref{tab:SW}.  In addition, Table \ref{tab:SWSR} records the average \emph{sparsity ratio} for each VARX-L model.  The sparsity ratio denotes the average proportion of model coefficients that are set to zero across the evaluation period.  A sparsity ratio of one implies that all coefficients are set to zero whereas a least squares fit that includes all coefficients has a sparsity ratio of zero. 
  \begin{table}[H]
    \centering    
 \caption{ \footnotesize One-step and four-step ahead MSFE of $k=20$ macroeconomic indicators (relative to sample mean) with $m=20$ exogenous predictors $p=4,s=4$.}
\label{tab:SW}
\footnotesize
\begin{tabular}{ l | c c| c c}
\hline 
Model/VARX-L Penalty Structure  & One-step ahead Out of Sample Relative MSFE &  Four-step ahead Out of Sample Relative MSFE\\
    \hline
Basic  &  0.8064 &  0.9672  \\
Lag Group  &0.8747   & 0.9798  \\
Own/Other Group   &  0.7773  & 0.9582  \\
Sparse Lag Group  & 0.8206 & 0.9702  \\
Sparse Own/Other Group  &   0.7823   & 0.9590 \\
Endogenous-First   &  0.8531 & 0.9748  \\ 
\hline
VARX with lags selected by AIC&  5.0223  &7.8363 \\
VARX with lags selected by BIC&  0.9455 & 1.1603 \\
BGR's Bayesian VAR & 0.9414  &0.9765  \\
Factor Model & 0.9904 & 0.9819\\
\hline
Sample Mean  & 1.0000& 1.0000  \\
Random Walk &  1.9909 &1.8706      
\end{tabular}
 \end{table}

  \begin{table}[H]
    \centering    
 \caption{ \footnotesize Average sparsity ratio (proportion of least squares coefficients set to zero) of VARX-L models in forecasting $k=20$ macroeconomic indicators with $m=20$ exogenous predictors $p=4,s=4$, $k^2p+kms=3,200$}
\label{tab:SWSR}
\footnotesize
\begin{tabular}{ l | c c| c c}
\hline 
VARX-L Penalty Structure  & One-step ahead Average Sparsity Ratio & Four-step ahead Average Sparsity Ratio   \\
    \hline
Basic  &  0.9090  &0.9636 \\
Lag Group  &0.2259   &0.6315  \\
Own/Other Group   & 0.3049 &0.7030  \\
Sparse Lag Group  &  0.6743 & 0.7095 \\
Sparse Own/Other Group &0.7118  &  0.8853 \\
Endogenous-First  &0.1411  & 0.6068 \\ 
\end{tabular}
 \end{table}

 Most of our VARX-L procedures substantially outperform the benchmarks at both forecast horizons, with the Own/Other Group VARX-L and Sparse Own/Other Group VARX-L achieving the best performance.  This provides evidence that making the distinction between a series' own lags and those of other series can improve forecasts in macroeconomic applications.  The relatively poor performance of the Lag Group VARX-L suggests that a lag based grouping may be too restrictive for such a disparate group of series and hence not appropriate for this application.

The imposition of sparsity appears to be crucial, as BGR's Bayesian VAR performs worse than all VARX-L procedures at both horizons except for the Lag Group VARX-L at $h=4$.  It performs very similarly to the least squares VARX with lags selected by BIC at $h=1$, but slightly better at $h=4$.  The factor model also performs very poorly at both forecast horizons, providing evidence that a low-rank structure is not appropriate for this application.

 The VARX with lags selected by AIC is substantially outperformed by the sample mean at both horizons, whereas the VARX with lags selected by BIC slightly outperforms it at $h=1$, but is outperformed at $h=4$.  Since AIC imposes a weaker penalty for higher lag orders than BIC, it has a tendency to construct overparameterized models, whereas BIC has a tendency to underfit and misses out on potential dynamic relationships that the VARX-L procedures are able to capture.  Since neither approach imposes variable selection, these models tend to result in very noisy multi-step ahead forecasts.

We observe that Basic VARX-L imposes the most sparsity at both horizons, which is expected as it is not bound to any group structure.  The Sparse Own/Other VARX-L, which has a very flexible group structure as well as the ability to impose within-group sparsity returns the second most sparse model across both horizons.   The Lag Group VARX-L, Own/Other Group VARX-L and the Endogenous-First VARX-L, all lacking the ability to impose within-group sparsity, return relatively less sparse solutions at $h=1$, but impose much more sparsity at $h=4$, which could reflect the decreasing predictive power of the data in forecasting longer horizons.

As an additional application to showcase the tractability of our methods on high-dimensional problems, we attempt to forecast the \emph{medium-large} (k=40) set of indicators using the remaining variables in the \emph{large} (m=128) set as exogenous predictors.
The forecast performance is recorded in Table \ref{tab:SW2} and the average sparsity ratios of the VARX-L procedures are recorded in Table \ref{tab:SWSR2a}.

  \begin{table}[H]
    \centering    
 \caption{ \footnotesize One-step ahead MSFE and four-step ahead MSFE of $k=40$ macroeconomic indicators (relative to sample mean) with $m=128$ exogenous predictors $p=4,s=4$.}
\label{tab:SW2}
\footnotesize
\begin{tabular}{ l | c c}
\hline 
Model/VARX-L Penalty Structure  & One-step ahead Out of Sample Relative MSFE & Four-step ahead Out of Sample Relative MSFE \\
    \hline
Basic  &  0.7810 & 0.9824 \\
Lag Group  &0.8550&  0.9756  \\
Own/Other Group   & 0.7587&  0.9680  \\
Sparse Lag Group  & 0.7899 &0.9575   \\
Sparse Own/Other Group  &   0.7917&0.9952 \\
Endogenous-First   &  0.8278 &  0.9657\\ 
\hline
VARX with lags selected by AIC&  1.9789 & 2.5105 \\
VARX with lags selected by BIC&  0.9999 & 0.9999\\
BGR's Bayesian VAR & 0.9062 &0.9646   \\
Factor Model & 0.9758 &1.0166 \\
\hline
Sample Mean  & 1.0000 & 1.0000  \\
Random Walk &  1.6763 & 1.6763      
\end{tabular}
 \end{table}

  \begin{table}[H]
    \centering    
 \caption{ \footnotesize Average sparsity ratio (proportion of least squares coefficients set to zero) of VARX-L models in forecasting $k=40$ macroeconomic indicators with $m=128$ exogenous predictors $p=4,s=4$, $k^2p+kms=26,880$}
\label{tab:SWSR2a}
\footnotesize
\begin{tabular}{ l | c c| c c}
\hline 
VARX-L Penalty Structure  & One-step ahead Average Sparsity Ratio & Four-step ahead Average Sparsity Ratio   \\
    \hline
Basic  &  0.9683 & 0.9862   \\
Lag Group  &0.6300 & 0.9584    \\
Own/Other Group   & 0.8144 &0.9257   \\
Sparse Lag Group  &  0.8577 & 0.8711 \\
Sparse Own/Other Group &0.9912 & 0.9825  \\
Endogenous-First  &0.0702 & 0.1697   \\ 
\end{tabular}
 \end{table}

In this application, we observe that the Own/Other Group VARX-L achieves the best forecasting performance at $h=1$, followed by the Basic VARX-L.  At $h=4$, the Sparse Lag Group VARX-L achieves the best performance.  At $h=1$, as in the previous application, all VARX-L procedures outperform the benchmark models.  However, at $h=4$, BGR's Bayesian VAR achieves the second best forecasting performance.

Most of our VARX-L methods impose considerably more sparsity in this larger application as compared to Table \ref{tab:SWSR}, which is to be expected as many of the included exogenous series are likely either irrelevant or redundant for forecasting purposes.  Strangely, the Endogenous-First VARX-L imposes less sparsity than in the previous application, but this could be the result of its unique nested penalty structure which forces endogenous coefficients to be active along with their exogenous counterparts.  It is possible that the relatively poor performance of several VARX-L models at $h=4$ is the result of not imposing enough sparsity.  For example, the Sparse Own/Other Group VARX-L imposes less sparsity at $h=4$ than $h=1$.

\subsubsection*{Canadian Macroeconomic Data Application}
We next consider a low-dimensional application in which we forecast Canadian indicators
 using US macroeconomic series as exogenous predictors.  As a small, relatively open economy Canada's macroeconomic indicators have been shown to be very sensitive to their US counterparts.  In particular, \cite{Racette} and \cite{cushman} demonstrate that the US Gross Domestic Product and Federal Funds Rate are very influential in modeling Canada's analogous monetary policy proxy variables.  Taking this into consideration, we forecast $k=4$ Canadian macroeconomic series using our previously defined \emph{medium} dataset as exogenous predictors ($m=20$).  The endogenous series are Canadian M1 (a measure of the liquid components of money supply), Canadian Industrial Production, Canadian GDP (relative to 2000), and the Canada/US Exchange Rate.

The Canadian series range from Quarter 3 of 1960 to Quarter 3 of 2007.  Quarter 3 of 1977 to Quarter 2 of 1993 is used for penalty parameter selection while Quarter 3 of 1993 to Quarter 3  of 2007 is used for forecast evaluation ($T=191$).  In addition to the standard benchmarks, we also compare against our procedures in the VAR-L framework, in which the exogenous predictors are ignored.  Our results are summarized in Tables \ref{tab:CA} and \ref{tab:CA2}.     
 
 \begin{table}[H]
    \centering    
\footnotesize
 \caption{ \footnotesize One-step ahead and four-step ahead MSFE (relative to sample mean) for VARX forecasts of $k=4$ Canadian macroeconomic indicators with $m=20$ exogenous predictors $p=4,s=4$ and VAR forecasts of 4 Canadian macroeconomic indicators, $p=4$.}
\label{tab:CA}
\begin{tabular}{ l | c | c }
\hline 
 Model/VARX-L Penalty Structure &  One-step ahead Out of Sample RMSFE & Four-step ahead Out of Sample RMSFE  \\
    \hline
Basic  &  0.8406 & 0.9187   \\
Lag Group & 0.8357 & 0.9285
\\
Own/Other Group & 0.8376 & 0.9143
\\
Sparse Lag Group& 0.8274 & 0.9129
\\
Sparse Own/Other Group  &0.8327 & 0.9181
\\
Endogenous-First  &  0.8454 & 0.9593 \\
\hline
VARX with lags selected by AIC  &1.3680 & 1.7739   \\
VARX with lags selected by BIC & 0.8785 & 1.0941  \\
BGR's Bayesian VAR (with exogenous series)   & 1.0058 & 0.9748 \\
\hline
\hline
Model/VAR-L Penalty Structure  & One-step ahead Out of Sample RMSFE & Four-step ahead Out of Sample RMSFE\\
\hline
Basic  &    0.8465 & 0.9645 \\
 Lag Group      &  0.8575 & 0.9965 \\
Own/Other Group & 0.8491 & 0.9604 \\
Sparse Lag Group & 0.8506 & 0.9623 \\
Sparse Own/Other Group  &  0.8493 & 0.9655   \\
\hline
VAR with lag selected by AIC   &0.9190 & 1.1365  \\
VAR with lag selected by BIC  &0.8785 & 1.0941 \\
BGR's Bayesian VAR (without exogenous series) & 1.0066 & 0.9891\\
Factor Model & 0.9033 & 1.0152\\
\hline
\hline
Sample Mean & 1.0000 & 1.0000
\\
Random Walk  & 1.3388 & 1.7180

\end{tabular}
 \end{table}

 Even in this low dimension, we find that all of our models substantially outperform the AIC and BIC benchmarks across both forecast horizons, with the Sparse Lag Group VARX-L achieving superior performance at both horizons.  This low-dimensional example is better suited toward lag based groupings than our previous application.  Consequently, the relative forecasting performance of the Lag Group VARX-L and Sparse Lag Group VARX-L improve substantially.

In addition, we find that our methods are able to effectively leverage relevant information from the exogenous predictors, as every VARX-L procedure achieves better out of sample performance than its corresponding VAR-L.  Conversely, the information criterion based VARX approaches fail to outperform their VAR counterparts.  At $h=1$ and $h=4$, BIC produces identical forecast error in both the VAR and VARX setting, indicating that it never selects any exogenous series.

 BGR's Bayesian VAR performs poorly in this scenario, achieving similar forecast performance to the sample mean across both horizons, both with and without exogenous series, outperforming only the Lag Group VAR-L at $h=4$.  Its poor performance across both settings suggests that imposing sparsity is desirable even in low-dimensional applications.  The factor model also performs poorly, indicating that a low-rank structure may not be appropriate for this small-scale macroeconomic forecasting application.

 \begin{table}[H]
    \centering    
\footnotesize
 \caption{ \footnotesize Average sparsity ratio (proportion of least squares coefficients set to zero) of VARX-L models in forecasting $k=4$ macroeconomic indicators with $m=20$ exogenous predictors $p=4,s=4$, $k^2p+kms=384$ and VAR-L models lacking exogenous series $p=4, k^2p=64$.}
\label{tab:CA2}
\begin{tabular}{ l | c | c }
\hline 
 Model/VARX-L Penalty Structure &  One-step ahead Average Sparsity Ratio & Four-step Average Sparsity Ratio  \\
    \hline
Basic  &  0.7882 & 0.9091   \\
Lag Group & 0.5953 &0.7733
\\
Own/Other Group & 0.5909 & 0.8125
\\
Sparse Lag Group& 0.7282 &0.7979
\\
Sparse Own/Other Group  &0.6764 &0.8554
\\
Endogenous-First  &  0.0005 & 0.0236 \\
\hline
\hline
Model/VAR-L Penalty Structure  & One-step ahead Average Sparsity Ratio & Four-step ahead Average Sparsity Ratio\\
\hline
Basic  &    0.6151 & 0.7958 \\
 Lag Group      &  0.0000 & 0.8403  \\
Own/Other Group & 0.0002 & 0.6104 \\
Sparse Lag Group & 0.2114 &0.3114 \\
Sparse Own/Other Group  &  0.2494 & 0.7000   \\
\end{tabular}
 \end{table}

We observe that less sparsity is imposed in the VAR framework as opposed to the VARX.  At $h=1$, the Lag Group VAR-L imposes no sparsity and the Own/Other Group VAR-L imposes a negligible amount.  This relative lack of sparsity is likely the result of the VAR-L models already operating from a reduced model space as compared to their VARX-L counterparts.  However, under both frameworks, considerably more sparsity is imposed at $h=4$.  Across all models, we find that the level of sparsity generally increases with the number of potential coefficients as well as across longer forecast horizons.

\subsection{Evaluating Model Performance with Model Confidence Sets}
\label{secMCS}
In addition to simply evaluating model performance based on relative MSFE, we also consider applying the model confidence sets framework proposed by \cite{Hansen2012} which conducts a sequence of pairwise hypothesis tests in order to construct a set of ``superior models,'' within which the null hypothesis of \emph{equal predictive ability} cannot be rejected.  

This procedure starts by computing the sum of squared forecast errors (SSFE) for each candidate model over the evaluation period ($T_2+1$ through $T$).  The SSFE for model $i$ at time $t$ is defined as $\text{SSFE}_{i,t}=\|\mathbf{y}_t-\hat{\mathbf{y}}_t^{(i)}\|_F^2$.  After constructing the SSFE, at each time $t$, the MCS procedure computes the pairwise loss differential for each pair of models $i<j$: $d_{i,j,t}=\text{SSFE}_{i,t}-\text{SSFE}_{j,t}$.

Relative model performance is assessed according to the hypothesis of equal predictive ability, which tests whether pairwise loss averaged over time differs from zero across all model combinations.  In order to evaluate this hypothesis, we construct the test statistic
\begin{align}
\label{EPA1}
 v_{i,j}=\frac{\bar{d}_{ij}}{\sqrt{\widehat{\text{var}}(\bar{d}_{ij})}}, 
\end{align}
in which $\bar{d}_{ij}=\frac{1}{T-T_2}\sum_{t=T_2}^{T} d_{i,j,t},$ and $\widehat{\text{var}}(\bar{d}_{ij})$ is estimated according to a block bootstrap procedure.  The asymptotic distribution of this test statistic is non-standard, hence it is also estimated using a block bootstrap in a manner similar to the variance. 

The model confidence sets algorithm initializes by setting $M$ equal to all candidate models and iteratively testing for equal predictive ability using the aggregate test statistic $V_{R,M}=\max_{i,j\in M} |v_{ij}|$.  If equal predictive ability is rejected at a given confidence level $1-\alpha$, the worst performing model (i.e. the model with the largest average pairwise loss differential) is removed from $M$ and the procedure is repeated on the reduced subset of models.  The algorithm terminates once equal predictive ability cannot be rejected.  For more details on the MCS methodology, consult \cite{Hansen2012} and \cite{MCSPackage}. 

We utilize the {\tt R} package {\tt MCS} \citep{MCSPackage} to implement this procedure.  Following the package's default settings, we choose $\alpha=.15$ and perform 5000 bootstrap replications.  Our resulting sets of equal predictive ability are displayed in Table \ref{tab:MCS}.    

 \begin{table}[H]
    \centering    
\footnotesize
 \caption{ Model Sets $M$ of Equal Predictive Ability ($\alpha=0.15$).  Within each set of models, we cannot reject the null hypothesis of equal predictive ability, though they achieve superior forecasting performance relative to all excluded models. }
\label{tab:MCS}
\begin{tabular}{ l | c |c  }
Application & One-Step $M$ & Four-Step $M$\\
\hline
\multirow{2}{*}{\cite{stockdataset} US Macroeconomic Data ($k=20$, $m=20)$} &Own/Other VARX-L & Own/Other VARX-L \\  & Sparse Own/Other VARX-L & Sparse Own/Other VARX-L \\
\hline
\cite{stockdataset} US Macroeconomic Data ($k=40$, $m=128)$ & Own/Other VARX-L & Sparse Lag Group VARX-L \\
& & Own/Other VARX-L\\
& & Endogenous-First VARX-L\\
& & BGR's Bayesian VAR\\
\hline
\multirow{4}{*}{Canadian Macroeconomic Data ($k=4$, $m=20$)} &Sparse Own/Other VARX-L & Sparse Own/Other VARX-L\\ &  Sparse Lag Group VARX-L & Basic VARX-L\\ &  & Sparse Lag Group VARX-L\\ & & Own/Other VARX-L 
\end{tabular}
 \end{table}

We find that the MCS procedure is able to not only distinguish between the forecasting performance of our VARX-L models and the benchmarks, but also within the VARX-L class of models.  In the Canadian macroeconomic data application, no VAR-L models are included at either forecast horizon, indicating that the exogenous series have predictive power.  We additionally find that either the Sparse Own/Other Group VARX-L or Own/Other Group VARX-L are in the MCS in every application.
This provides further evidence supporting the use of a group structure that distinguishes between a series' own lags and those of other series in macroeconomic applications.  The only application in which a competitor's model is included in the MCS is the large US macroeconomic application ($k=40, m=128$) at $h=4$, though several VARX-L models are also included.  The relatively poor performance of the VARX-L models in this scenario suggests that long-horizon forecasts in large models deserve additional scrutiny.

\section{Extending the VARX-L for Unit-Root Nonstationarity}
\label{sec8}
In some scenarios, it may not be appropriate to shrink every coefficient toward zero.  In traditional time series analysis, economic series that exhibit persistence are transformed to stationarity.  However, this framework has several drawbacks.  First, if no pre-established transformation guidelines are available, this process can be labor intensive and subjective.  Second, as stated by \cite{Kennedy}, stationarity transformations destroy information about the long-run relationships of economic variables.  Ideally, to effectively forecast using all available information, it would be preferable to work directly with the untransformed series.  In this section, we outline a possible extension that allows for shrinking toward reference models, such as a vector random walk, that can account for mild non-stationarity, which is ubiquitous in macroeconomic data.
 \subsection*{The ``Minnesota'' VARX-L}

The proposed VARX-L models can easily be modified to shrink coefficients toward a known constant matrix.  Shrinking toward constant matrices $\mathbf{C}_y\in \mathbf{R}^{k\times kp}, \mathbf{C}_x \in \mathbf{R}^{k\times ms}$ results in a slightly modified objective of the form  
\begin{align}
\label{MN1}
\min_{\utwi{\nu},\PhiB,\betaB}  \sum_{t=1}^T\|\mathbf{y}_t-\utwi{\nu}-\PhiB\Y_{t-1}-\betaB\mathbf{X}_{t-1}\|_F^2+\lambda \bigg(\mathcal{P}_y(\PhiB-\mathbf{C}_y)+ \mathcal{P}_x(\betaB-\mathbf{C}_x)\bigg).
\end{align}
in which $\Y_t=
[\mathbf{y}_t^{\top},\dots,\mathbf{y}_{t-p}^{\top}]$ and $\mathbf{X}_t=[\mathbf{x}_{t}^{\top},\dots,\mathbf{x}_{t-s}^{\top}]$.

Let $[\PhiB,\betaB]^{\lambda}(\mathbf{C}_y,\mathbf{C}_x)$
denote a solution to this problem.  Now, by a change of variables $\widetilde{\PhiB}=\PhiB-\mathbf{C}_y$ and $\widetilde{\betaB}=\betaB-\mathbf{C}_x$, we obtain the equivalent problem
\begin{align*}
 \min_{\utwi{\nu},\widetilde{\PhiB},\widetilde{\betaB}} \sum_{t=1}^T\|\mathbf{y}_t-\utwi{\nu}-\mathbf{C}_y\Y_{t-1}-\widetilde{\PhiB}\Y_{t-1}-\mathbf{C}_x\mathbf{X}_{t-1}-\widetilde{\betaB}\mathbf{X}_{t-1}\|_F^2+\lambda \bigg(\mathcal{P}_y(\widetilde{\PhiB})+ \mathcal{P}_x(\widetilde{\betaB})\bigg),
\end{align*}
which can be expressed as
\begin{align*}
 &\min_{\utwi{\nu},\widetilde{\PhiB},\widetilde{\betaB}} \sum_{t=1}^T\|\widetilde{\mathbf{y}}_t-\utwi{\nu}-\widetilde{\PhiB}\Y_{t-1}-\widetilde{\betaB}\mathbf{X}_{t-1}\|_F^2+\lambda \bigg(\mathcal{P}_y(\widetilde{\PhiB})+ \mathcal{P}_x(\widetilde{\betaB})\bigg),
\end{align*}
in which $\widetilde{\mathbf{y}}_t=\mathbf{y}_t-\mathbf{C}_y\Y_{t-1}-\mathbf{C}_x\mathbf{X}_{t-1}$.
We can view the solution to this transformed problem as $[\widetilde{\PhiB},\widetilde{\betaB}]^{\lambda}(\utwi{0},\utwi{0})$ operating on $\widetilde{\mathbf{y}}_t$.  Hence, transforming back to the setting of Equation \eqref{MN1}, we find that
\begin{align*}
&[\PhiB,\betaB]^{\lambda}(\mathbf{C}_y,\mathbf{C}_x)=[\mathbf{C}_y,\mathbf{C}_x]+[\widetilde{\PhiB},\widetilde{\betaB}]^{\lambda}(\utwi{0}_{k\times kp},\utwi{0}_{k\times ms}).
\end{align*}

As an example, consider $\mathbf{C}_y=[\mathbf{I}_{k},\mathbf{0}_{k\times k},\dots,\mathbf{0}_{k\times k}],\mathbf{C}_x=\mathbf{0}_{k\times ms}$, which implements a variant of the Minnesota prior, shrinking the VARX-L model toward a vector random walk.  We refer to this extension as the ``Minnesota'' VARX-L.  It could be very useful in economic applications as it is widely believed that many persistent macroeconomic time series can be well approximated by a random walk \citep{Litterman1979}.

In order to validate this alternative approach, we follow the methodology of \cite{BGR}, who also utilize the data from \cite{stockdataset}, but eschew stationarity transformations and work directly with the untransformed series.  We again apply our VARX-L forecasting procedures by forecasting the aforementioned \emph{medium} set of $(k=20)$ series using the remaining 20 variables in the \emph{medium large} set as exogenous predictors, but choose not to perform any stationarity transformations and instead shrink toward a vector random walk.

One advantage of not applying stationarity transformations is that it allows us to utilize more of our data.  The data used in section \ref{sec53} extends to Quarter 4 of 2008, but one series, non-borrowed depository institutional reserves (FMRNBA), becomes negative in early 2008 due in part to changes in both monetary policy and Federal Reserve accounting \citep{WSJarticle}.  The stationarity transformation guidelines provided by \cite{stockdataset} for this series propose taking the first difference of logs, which is obviously not appropriate for negative values.

Quarter 3 of 1976 to Quarter 2 of 1993 are used for penalty parameter selection while Quarter 3 of 1993 through Quarter 4 of 2008 are used for forecast evaluation (T=200). In this application, we also shrink BGR's Bayesian VAR toward a random walk.  Our results are summarized in Table \ref{tab:tabSWRW} and the average sparsity ratios are recorded in Table \ref{tab:SWSRNS}.

  \begin{table}
    \centering    
\footnotesize
 \caption{ \footnotesize
 One-step and four-step ahead MSFE (relative to a random walk) for $k=20$ nonstationary macroeconomic indicators with m=20 exogenous predictors which shrink toward a vector random walk.}
\label{tab:tabSWRW}
   \begin{tabular}{ l | c |c }
\hline 

Model/Minnesota VARX-L Penalty Structure & One-step Ahead Out of Sample RMSFE &  Four-step Ahead Out of Sample RMSFE \\
    \hline
Basic  & 0.8173 & 0.9460 \\
Lag Group     &  0.9450 & 0.9590 \\
Own/Other Group  &0.8155 & 0.9520   \\
Sparse Lag Group      & 0.9858 & 0.9702   \\
Sparse Own/Other Group    & 0.8808 & 0.9550     \\
Endogenous-First      & 0.9746 & 0.9518     \\
\hline
VARX with lag selected by AIC   & 1.2764 & 1.1896 \\
VARX with lag selected by BIC   & 1.2764 & 1.1896\\
BGR's Bayesian VAR  & 1.3475 & 1.0083 \\
Factor Model & 4.7979 & 1.5968\\
Sample Mean   & 11.304 & 5.7747 \\
Random Walk  & 1.0000 & 1.0000
  \end{tabular}
 \end{table}

  \begin{table}[H]
    \centering    
 \caption{ \footnotesize Average sparsity ratios (proportion of least squares coefficients set to zero) of VARX-L models in forecasting $k=20$ nonstationary macroeconomic indicators with $m=20$ exogenous predictors $p=4,s=4$, $k^2p+kms=3200$}
\label{tab:SWSRNS}
\footnotesize
\begin{tabular}{ l | c c| c c}
\hline 
VARX-L Penalty Structure  & One-step ahead Average Sparsity Ratio & Four-step ahead Average Sparsity Ratio   \\
    \hline
Basic  &  0.9746  &0.9652 \\
Lag Group  &0.7650   &0.7595  \\
Own/Other Group   & 0.7602 &0.7533  \\
Sparse Lag Group  &  0.8484 & 0.8842 \\
Sparse Own/Other Group &0.8004  & 0.8099 \\
Endogenous-First  &0.7594  & 0.7875 \\ 
\end{tabular}
 \end{table}

 We find the each of these Minnesota VARX-L procedures outperform the random walk at both forecast horizons with the Own/Other Group Minnesota VARX-L achieving the best out of sample performance at $h=1$ and the Basic VARX-L performing the best at $h=4$.  

 We observe that under this scenario, the choice of structure substantially affects forecasting performance.  Lag based groupings, such as the Lag Group, Sparse Lag Group, and Endogenous-First perform relatively poorly at $h=1$ but slightly improve relative to other methods at $h=4$, however they still outperform the naive methods across both horizons.  Their reduced relative performance is likely due to their inability to distinguish between the diagonal random walk component and the coefficients on other lags in the lag one coefficient matrix $\PhiB^{(1)}$.   

 AIC and BIC are not well suited toward a nonstationarity setting, hence are completely uninformative, selecting lag orders of $p=1$ and $s=0$ at every point in time across both horizons.  BGR's procedure, despite the imposition of a random walk prior, produces inferior forecasts to both the VARX-L procedures and the naive random walk.  The factor model is also poorly suited for this application and is substantially outperformed by the random walk.

A considerable amount of sparsity is imposed across forecast horizons, regardless of structure with the Basic VARX-L returning the most sparse models.  This indicates that a greater level of sparsity may be more appropriate when shrinking toward a reference model as opposed to identically toward zero.

\section{Simulation Scenarios}
\label{sec52}
In this section, we consider evaluating the forecasting performance of our procedures on several simulated multivariate time series conforming to different sparsity patterns, with one constructed to be advantageous for each proposed structure.  Note that since the factor model is designed to forecast well in low-rank regimes as opposed to sparse regimes, due to its anticipated poor performance, we omit it from this section.    

Our objective is to quantify relative performance under both matched and unmatched model sparsity and penalty function structures.  All simulations operate on a $\text{VARX}_{5,5}(4,4)$ of length $T=100$, and each simulation is repeated 100 times.  The choice of $p=s=4$ was selected because it represents one year of dependence for quarterly series, which is a common frequency of macroeconomic data.  The middle third of the data is used for penalty parameter selection while the last third is used for forecast evaluation.  Under the first 5 scenarios, $\Sigma_u$ is distributed according to a multivariate normal distribution with mean $\utwi{0}_{5}$ and covariance $(0.01)\times\utwi{I}_{5}$; the sixth scenario utilizes a more general specification.  We do not include an intercept in any simulation scenarios.  The coefficient matrix from each simulation scenario was designed to ensure that a stationary process would be generated.  This procedure is elaborated upon in \cite{BigVAR}.    

In order to simulate from a $\text{VARX}_{5,5}(4,4)$, we start by constructing a $\text{VAR}_{10}(4)$.  Denoting the first 5 series as $\mathbf{y}_t$ and the second 5 as $\mathbf{x}_t$, we simulate according to the unidirectional relationship
\begin{align*}
  \begin{pmatrix}
   \mathbf{y}_{t}\\
   \mathbf{x}_t
  \end{pmatrix}
=
\mathlarger{\mathlarger{â€Ã¢€Ž\sum}}_{\ell=1}^{4}
\begin{pmatrix}
 \PhiB^{(\ell)} & \betaB^{(\ell)}\\ 
\utwi{0} & \mathbf{\Gamma}^{(\ell)}
\end{pmatrix}
  \begin{pmatrix}
   \mathbf{y}_{t-\ell}\\
   \mathbf{x}_{t-\ell}
  \end{pmatrix}
+\mathbf{u}_t,
\end{align*}
in which $\mathbf{\Gamma}^{\ell}\in \mathbb{R}^{m\times m}$ denotes the dependence structure of the exogenous series $\mathbf{x}_t$ (which follows the same sparsity pattern as $\PhiB^{\ell}$), and
$\mathbf{u}_t\stackrel{\text{iid}}{\sim}N(\utwi{0},\Sigma_u)$.  The residual covariance $\Sigma_u$ is set to $0.01\times I_{10}$ for the first 5 scenarios; the covariance used in Scenario 6 is depicted in Figure \ref{fig:cov}.

\subsubsection*{Scenario 1:  Unstructured Sparsity}
\begin{figure}
\centering
\footnotesize
\caption{ \footnotesize  \label{fig:figsp1a} Sparsity Pattern Scenario 1: Unstructured Sparsity.  Darker shading represents coefficients that are larger in magnitude.}
\includegraphics[scale=.90]{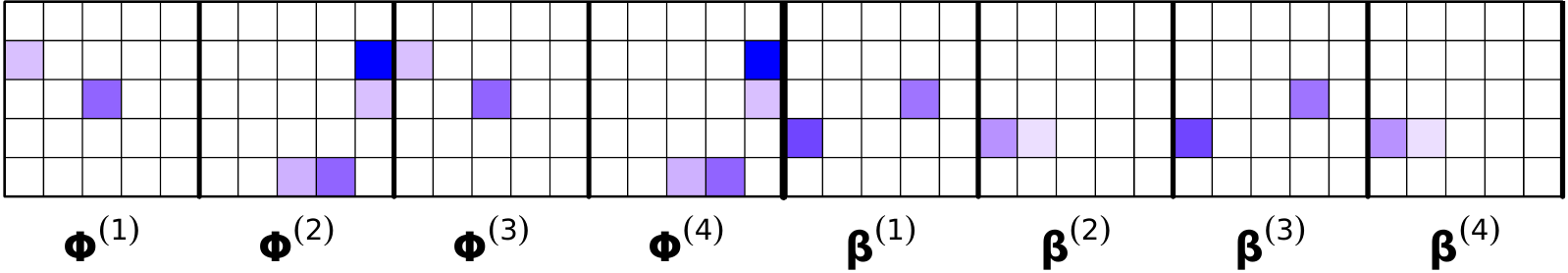}
\end{figure}

We first consider a scenario in which the sparsity is completely random; our sparsity pattern was generated in such a manner that each coefficient was given an equal probability (10 percent) of being active, resulting in a coefficient matrix in which roughly 90 percent of coefficients are zero.  Under such a design, we should expect superior performance from the Basic VARX-L, which assumes no group structure.  We do not expect such a structure to be a common occurrence in macroeconomic applications, but it may be present in other application areas, such as internet traffic in which the included series can differ substantially and will likely not exhibit any group structure.   This sparsity pattern is depicted in Figure \ref{fig:figsp1a} and the results are summarized in Table \ref{tab:tabres1}.

  \begin{table}
    \centering    
\footnotesize
 \caption{\footnotesize Out of sample MSFE of one-step ahead forecasts after 100 simulations: Scenario 1.  Standard errors are shown in parentheses.}
\label{tab:tabres1}
\begin{tabular}{ l | c | c }
\hline 
Model/VARX-L Penalty Structure  & MSFE & MSFE Relative to Sample Mean   \\
    \hline
Basic  &  0.0645 (0.0012) & 0.0454    \\
Lag Group & 0.0755 (0.0010) & 0.0532   \\
Own/Other Group  &  0.0734 (0.0010) & 0.0517 \\
Sparse Lag Group &0.0724 (0.0009) & 0.0510  \\
Sparse Own/Other Group
& 0.0699 (0.0009) & 0.0492    \\
Endogenous-First
& 0.0779 (0.0010) & 0.0549   \\
\hline 
VARX with lags selected by AIC
&0.1040 (0.0017) & 0.0733  \\
VARX with lags selected by BIC
&0.1183 (0.0032) & 0.0833   \\
BGR's Bayesian VAR  & 0.3675 (0.0124)  & 0.2590   \\
\hline
Sample Mean
& 1.4187 (0.0681) & 1.0000  \\
Random Walk
& 0.8416 (0.0272)   & 0.5932   
\end{tabular}
 \end{table}

In this scenario, as expected, we find that the Basic VARX-L achieves the best performance.  Of the structured methods, the Sparse Own/Other VARX-L performs the best, as it can partially accommodate this sparsity pattern.  As expected, the other approaches, which impose a structure that is not present in the data suffer from degraded forecasts, but all structured approaches substantially outperform the AIC and BIC benchmarks.  BGR's Bayesian VAR, which cannot perform variable nor lag order selection, achieves substantially worse forecast performance than both information criterion based methods.

\subsubsection*{Scenario 2: Lag Group Sparsity}
\begin{figure}
\centering
\footnotesize
\caption{ \footnotesize
  \label{fig:figsp2} Sparsity Pattern Scenario 2: Lag Group Sparsity}
\includegraphics[scale=.90]{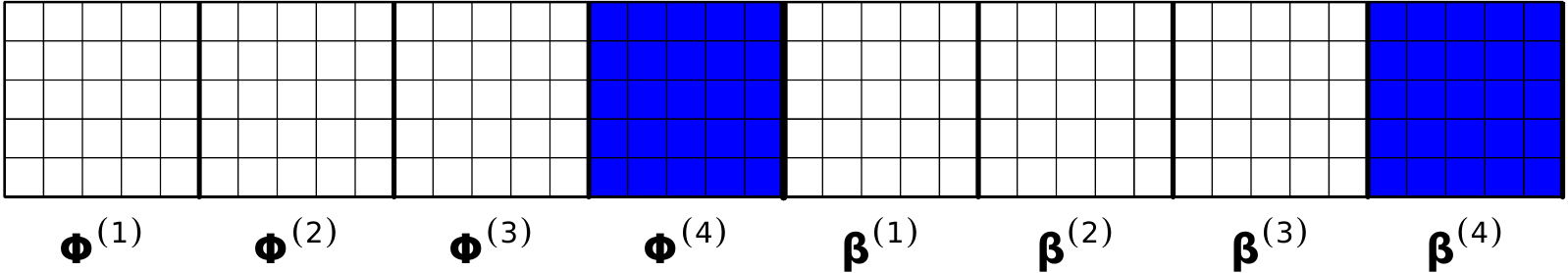}
\end{figure}

We next consider a scenario in which $\PhiB^{(4)}$ and $\betaB^{(4)}$ are dense with coefficients of the same magnitude, and all other coefficients are set to zero.  Such a sparsity pattern may be present in disaggregated macroeconomic series, such as agricultural price indices which follow a purely seasonal autoregressive relationship and exhibit a substantial degree of cross-dependence.  Under such a design, we should expect superior performance from the Lag Group VARX-L, which partitions all coefficients within a lag to the same group.  This sparsity pattern is depicted in Figure \ref{fig:figsp2}, and the results are summarized in Table \ref{tab:tabres2}.  

  \begin{table}
    \centering    
\footnotesize
 \caption{\footnotesize
   Out of sample MSFE of one-step ahead forecasts after 100 simulations: Scenario 2.  Standard errors are shown in parentheses.}
\label{tab:tabres2}
\begin{tabular}{ l | c | c}
\hline 
Model/VARX-L Penalty Structure & MSFE & MSFE Relative to Sample Mean    \\
    \hline
Basic  &    0.0786 (0.0012) &0.1397 \\
Lag Group & 0.0709 (0.0011)&0.1260  \\
Own/Other Group   &  0.0713 (0.0011) & 0.1268 \\
Sparse Lag Group &0.0739 (0.0012) & 0.1314  \\
Sparse Own/Other Group &0.0742 (0.0011) &0.1319     \\
Endogenous-First  & 0.0720 (0.0011) &0.1280 \\ 
\hline
VARX with lags selected by AIC  & 1.0084 (0.0273) & 1.7933  \\
VARX with lags selected by BIC  & 0.9927 (0.0282) & 1.7654  \\
BGR's Bayesian VAR  & 0.5769 (0.0146) & 1.0259   \\
\hline
Sample Mean  & 0.5623 (0.0123) & 1.0000  \\
Random Walk & 1.1279 (0.0322)  &2.0058    
\end{tabular}
 \end{table}

As expected, we find that the Lag Group VARX-L achieves the best performance and all structured approaches outperform the Basic VARX-L.  Under this scenario, all VARX-L procedures offer a substantial improvement over the benchmarks.  This is likely a result of their ability to effectively leverage the strong signal from the exogenous predictors.  Note that although the AIC and BIC benchmarks utilize this exogenous information, they are restricted to select from models of sequentially increasing lag order, hence they cannot accommodate this sparsity pattern and likely overfit, resulting in comparable performance to a random walk.  BGR's Bayesian VAR improves upon the information criterion based benchmarks, but since it cannot perform variable selection, it performs substantially worse than all VARX-L methods.

\subsubsection*{Scenario 3: Structured Lagwise Sparsity, Unstructured Within-Lag}
\begin{figure}
\centering
\caption{ \footnotesize  \label{fig:figsp3} Sparsity Pattern Scenario 3: Structured Lagwise, Unstructured within Lag}
\includegraphics[scale=.90]{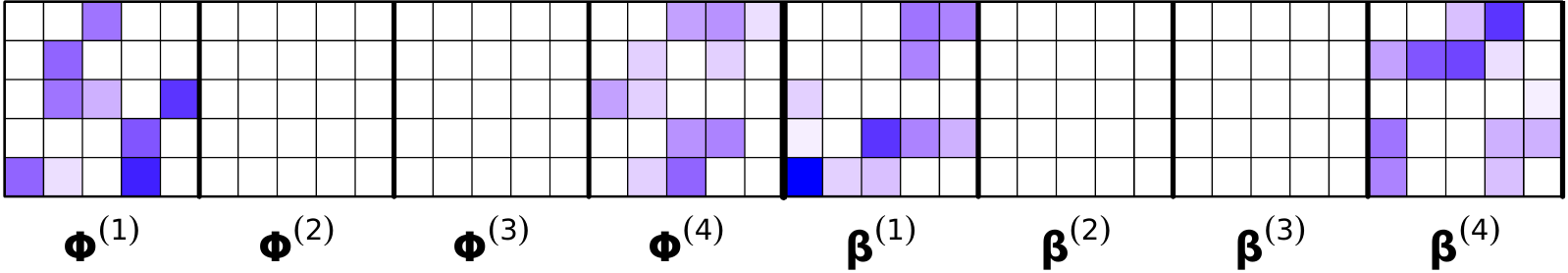}
\end{figure}

Our third scenario can be thought of as a hybrid of Scenarios 1 and 2.  As in Scenario 2, certain coefficient matrices are set identically to zero; only matrices $\PhiB^{(1)}$, $\PhiB^{(4)}, \betaB^{(1)},$ and $\betaB^{(4)}$ contain nonzero coefficients.  Additionally, in a similar manner to Scenario 1, sparsity within each lag is unstructured.  This scenario can be viewed as a less restrictive and more realistic version of the structure presented in Scenario 2 as it allows the degree of cross-dependence to vary across components.  In such a scenario, we should expect procedures that allow for within-group sparsity, such as the Sparse Lag Group VARX-L and Basic VARX-L to achieve the best forecast performance.   This sparsity pattern is depicted in Figure \ref{fig:figsp3} and the results are summarized in Table \ref{tab:tabres4}.

  \begin{table}
    \centering    
\footnotesize
 \caption{ \footnotesize Out of sample MSFE of one-step ahead forecasts after 100 simulations: Scenario 3.  Standard errors are shown in parentheses.}
\label{tab:tabres4}
\begin{tabular}{ l | c| c}
\hline 
Model/VARX-L Penalty Structure & MSFE & MSFE Relative to Sample Mean     \\
    \hline
Basic &    0.0665 (0.0008) & 0.1258 \\
Lag Group & 0.0696 (0.0008) &0.1317  \\
Own/Other Group  &  0.0699 (0.0009) & 0.1322 \\
Sparse Lag Group &0.0677 (0.0008) & 0.1281  \\
Sparse Own/Other Group  & 0.0683 (0.0008) & 0.1293  \\
Endogenous-First  & 0.0711 (0.0009) &0.1345 \\ 
\hline
VARX with lags selected by AIC  &0.1300 (0.0019) & 0.2458  \\
VARX with lags selected by BIC  &0.2501 (0.0061) & 0.4730 \\
BGR's Bayesian VAR  & 0.7568 (0.0515) & 1.4314   \\
\hline
Sample Mean  & 0.5287 (0.0275) & 1.0000  \\
Random Walk & 1.3000 (0.0731)  & 2.4588     
\end{tabular}
 \end{table}

Under this scenario, the Basic VARX-L achieves the best performance, followed closely by the Sparse Lag Group VARX-L.  Unlike Scenario 2, since this structure exhibits dependence in the first lag, the information-criterion based benchmarks are able to capture a portion of the true underlying structure in both the endogenous and exogenous series and thus substantially outperform the naive benchmarks.  However, since they cannot account for within-lag sparsity, they are still considerably outperformed by all VARX-L methods.  As in Scenario 1, BGR's Bayesian VAR performs very poorly, since it cannot perform variable or lag order selection.

\subsubsection*{Scenario 4: Sparse and Diagonally Dominant}
\begin{figure}
\centering
\caption{ \footnotesize \label{fig:figsp1} Sparsity Pattern Scenario 4: Sparse and Diagonally Dominant}
\includegraphics[scale=.90]{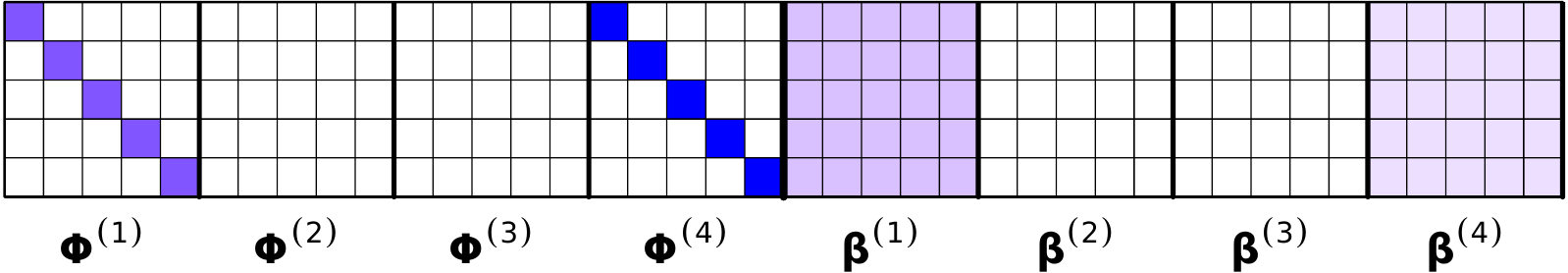}
\end{figure}

Our final scenario consists of a diagonally-dominant sparsity structure, in which all diagonal elements in $\PhiB^{(1)}$ and $\PhiB^{(4)}$ are equal in magnitude, whereas all off-diagonal endogenous coefficients are set to zero.  As in scenario 2, the coefficients in $\betaB^{(1)}$ and $\betaB^{(4)}$ are identical in magnitude.  This structure incorporates the belief posited by \cite{Litterman5yr} that macroeconomic series' own lags are more informative in forecasting applications than lags of other series.  Under this setting, one would expect superior performance from the Own/Other Group VARX-L.  The sparsity pattern is depicted in Figure \ref{fig:figsp1} and the simulation results are summarized in Table \ref{tab:tabres3}.

  \begin{table}
    \centering    
\footnotesize
  \caption{\footnotesize Out of sample MSFE of one-step ahead forecasts after 100 simulations: Scenario 4.  Standard errors are shown in parentheses.}
\label{tab:tabres3}
\begin{tabular}{ l |c |c }
\hline 
Model/VARX-L Penalty Structure  & MSFE  & MSFE Relative to Sample Mean  \\
    \hline
Basic   &    0.0669 (0.0008)& 0.0406  \\
Lag Group &0.0720 (0.0008) & 0.0437 \\
Own/Other Group   &  0.0626 (0.0008)& 0.0380 \\
Sparse Lag Group &0.0729 (0.0011) & 0.0442 \\
Sparse Own/Other Group  & 0.0625 (0.0008) & 0.0379  \\
Endogenous-First  & 0.0725 (0.0011) & 0.0440\\ 
\hline
VARX with lags selected by AIC  & 0.1043 (0.0015) & 0.0633 \\
VARX with lags selected by BIC  &0.1044 (0.0015) &0.0634 \\
BGR's Bayesian VAR  & 0.7741 (0.0394) &0.4702  \\

\hline
Sample Mean  & 1.6460 (0.0902) &1.0000 \\
Random Walk & 0.7512 (0.0390) & 0.4563    
\end{tabular}
 \end{table}

Under Scenario 4, as expected, the Own/Other Group and Sparse Own/Other Group VARX-L achieve superior forecasts.  Since the magnitude of coefficients within a lag matrix varies substantially, structures that utilize lag-based groupings, such as the Lag Group and Endogenous-First VARX-L are unable to capture this discrepancy and thus perform relatively poorly.  However, they still substantially outperform the benchmark procedures.  We again find that VARX with lags selected by AIC and BIC perform very poorly, as they are restricted to select from sequentially increasing lag orders and cannot account for within-lag sparsity.  BGR's Bayesian VAR also performs poorly for similar reasons.

\subsubsection*{Scenario 5: No within-lag Sparsity}
\begin{figure}
\centering
\caption{ \footnotesize \label{fig:figsp5} Magnitudes of Coefficients Scenario 5: No Sparsity}
\includegraphics[scale=.90]{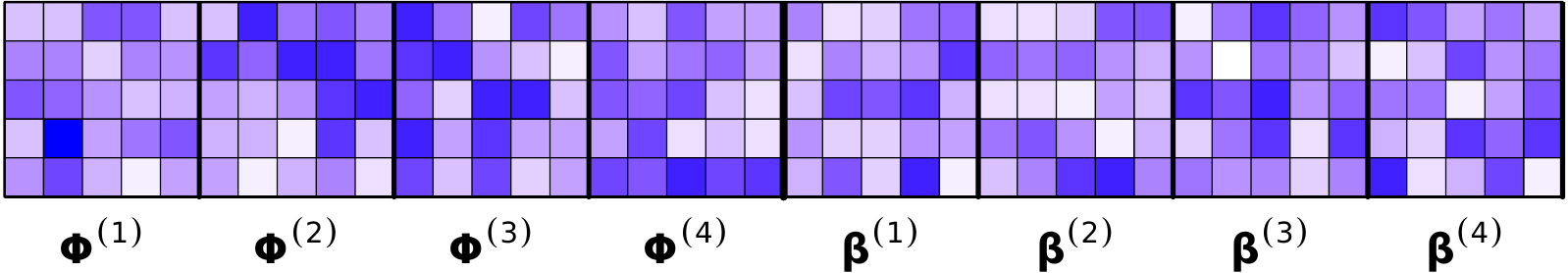}
\end{figure}

Next, we consider a scenario in which the entire $5\times 40$ coefficient matrix is dense.  Although the coefficient matrix is not sparse, as in the other scenarios, the information criterion based methods can impose sparsity by truncating the maximum lag orders $\hat{p}$ and $\hat{s}$.  The relative magnitudes of the coefficient matrix are depicted in Figure \ref{fig:figsp5}.  This scenario serves to give an advantage to the procedures that do not impose any within-lag sparsity: the information criterion based methods as well as BGR's Bayesian VAR.  The results from this scenario are depicted in Table \ref{tab:tabres5}.

  \begin{table}
    \centering    
\footnotesize
 \caption{\footnotesize Out of sample MSFE of one-step ahead forecasts after 100 simulations: Scenario 5.  Standard errors are shown in parentheses.}
\label{tab:tabres5}
\begin{tabular}{ l | c | c }
\hline 
Model/VARX-L Penalty Structure  & MSFE & MSFE Relative to Sample Mean   \\
    \hline
Basic  &  0.0780 (0.0010) & 0.6142    \\
Lag Group & 0.0731 (0.0008) & 0.5724   \\
Own/Other Group  &  0.0736 (0.0009) & 0.5761 \\
Sparse Lag Group &0.0747 (0.0009) & 0.5848  \\
Sparse Own/Other Group
& 0.0747 (0.0009) & 0.5849    \\
Endogenous-First
&0.0736  (0.0009) & 0.5766   \\
\hline 
VARX with lags selected by AIC
&0.1207 (0.0016) & 0.9444  \\
VARX with lags selected by BIC
&0.1157 (0.0022) & 0.9053   \\
BGR's Bayesian VAR  &  0.1148 (0.0018)  & 0.8982   \\
\hline
Sample Mean
&0.1278 (0.00296)  & 1.0000  \\
Random Walk
& 0.2020 (0.00369)   & 1.5806   
\end{tabular}
 \end{table}
We observe that despite a lack of sparsity in the data generating process, all VARX-L methods substantially outperform the information criterion benchmarks as well as BGR's Bayesian VAR.  The Lag Group VARX-L achieves the best performance, which is to be expected.  Since the Lag Group VARX-L has fewer groups than the Own/Other structures and does not impose within-group sparsity, it has more of a tendency to employ ridge-like penalization as opposed to setting an entire group to zero.  Within the class of VARX-L models, the Basic VARX-L performs the worst.  This suggests that structured groupings are more robust in applications where the true model is not sparse.

\subsubsection*{Scenario 6: Non-diagonal Covariance}
\begin{figure}
\centering
\caption{ \footnotesize \label{fig:cov} Covariance Matrix: Simulation Scenario 6}
\includegraphics[scale=.25]{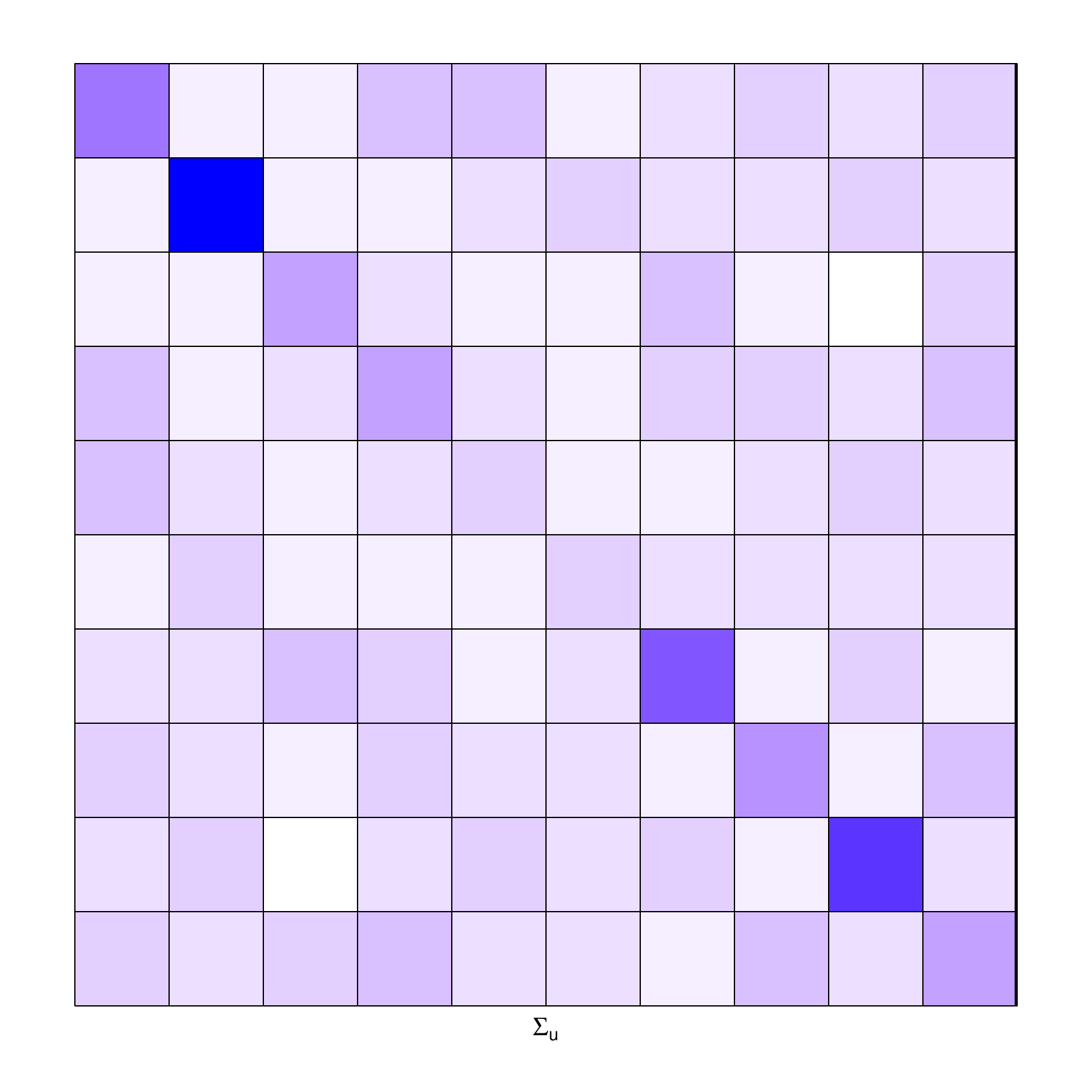}
\end{figure}

The previous five simulation scenarios impose a diagonal structure on $\Sigma_u$, the contemporaneous covariance matrix.  Such a scenario may rarely occur in practice.  In order to examine the robustness of the VARX-L procedures in the presence of a non scaled-identity covariance matrix, this scenario pairs the sparsity pattern from Scenario 1 with the covariance structure depicted in Figure \ref{fig:cov}.

Note that unlike the least squares VARX with lags selected by AIC or BIC, as well as BGR's Bayesian VAR, the VARX-L procedures do not explicitly incorporate $\Sigma_u$.  Hence, one should expect the benchmark procedures to be at a slight advantage in this scenario.  The results from this scenario are summarized in Table \ref{tab:tabres6a}.

  \begin{table}
    \centering    
\footnotesize
 \caption{\footnotesize Out of sample MSFE of one-step ahead forecasts after 100 simulations: Scenario 6.  Standard errors are shown in parentheses.}
\label{tab:tabres6a}
\begin{tabular}{ l | c | c }
\hline 
Model/VARX-L Penalty Structure  & MSFE & MSFE Relative to Sample Mean   \\
    \hline
Basic  &  0.7982 (0.0113) & 0.1034    \\
Lag Group & 0.9089 (0.0133) & 0.1177   \\
Own/Other Group  &  0.8926 (0.0134) & 0.1156 \\
Sparse Lag Group &0.8833 (0.0122) & 0.1144  \\
Sparse Own/Other Group
& 0.8746 (0.0127) & 0.1133    \\
Endogenous-First
&0.9283  (0.0135) & 0.1202   \\
\hline 
VARX with lags selected by AIC
&1.4237 (0.0232) & 0.1844  \\
VARX with lags selected by BIC
&1.0895 (0.0185) & 0.1411   \\
BGR's Bayesian VAR  & 2.8876 (0.0782)  & 0.4022   \\
\hline
Sample Mean
&7.1789 (0.3739)  & 1.0000  \\
Random Walk
& 4.2543 (0.1465)   & 0.5511   
\end{tabular}
 \end{table}
In this scenario (see Table \ref{tab:tabres6a}) we observe that the Basic VARX-L, which conforms to the true sparsity pattern achieves the best performance, following by the Sparse Own/Other Group VARX-L despite the more complex error structure.  As in every other scenario, all VARX-L models substantially outperform the benchmarks.  BGR's Bayesian VAR performs substantially worse than both information criterion based methods, even though it explicitly incorporates the covariance of $\mathbf{u}_t$; its poor performance is likely the result of incorporating an unreliable covariance estimate.   For an expanded discussion and simulation study involving non-diagonal covariance structures as well as a procedure to incorporate covariance in fitting VARX-L models, consult \cite{BigVAR}. 

Overall, all of the proposed VARX-L models are fairly robust to sparsity patterns not conforming to their true group structures.  In each scenario, every method substantially outperforms all benchmark procedures.  Scenario 1 is the only case in which the structured approaches perform poorly relative to the Basic VARX-L.  We expect such an unstructured sparsity pattern to occur only rarely in macroeconomic applications.

\section{Conclusion}
\label{sec7}

We have shown that the proposed VARX-L structured regularization framework is very amenable to the VARX setting in that it can simultaneously reduce its parameter space and still incorporate useful information from both endogenous and exogenous predictors.  VARX-L models scale well with the dimension of the data and are quite flexible in accommodating a wide variety of potential dynamic structures.  Each of the proposed methods consistently outperforms benchmark procedures both in simulations and in macroeconomic forecasting applications.  Forecast performance of all models appears to be robust across multiple sparsity structures as well as forecast horizons.  Moreover, upon examining actual macroeconomic data, structured VARX-L models tend to outperform the Basic VARX-L.  

Our work has considerable room for extensions.  This paper focuses solely on forecasting applications, but our VARX-L framework could also be extended to structural analysis and policy evaluation using an approach similar to that of \cite{furman}.  In addition, our current implementation requires a coherent maximal lag selection mechanism.  The common procedure of choosing a lag order based on the frequency of the data is problematic in that it can lead to overfitting.  One could potentially incorporate an additional penalty parameter that grows as the lag order increases, as in \cite{BickelSong}, but this approach requires a multi-dimensional penalty parameter selection procedure and subjective specification of a functional form for the lag penalty.

An {\tt R} package containing our algorithms and validation procedures, {\tt  BigVAR}, is available on the Comprehensive R Archive Network (CRAN).

\section*{Acknowledgements}
The authors thank Gary Koop for providing his data transformation script, Marta Ba\'{n}bura, Domenico Giannone, and Lucrezia Reichlin for sharing their BVAR code, and the attendees of the 2014 NBER/NSF Time Series Conference for their constructive comments as well as the helpful suggestions of two anonymous referees.  This research was supported by an Amazon Web Services in Education Research Grant.  DSM was supported by a Xerox PARC Faculty Research Award and National Science Foundation Grant DMS-1455172.  JB was supported by National Science Foundation DMS-1405746.

\bibliography{finalbib}

\appendix
\section{Appendix}

\subsection{Compact Matrix Notation}

In deriving the solution methods for our algorithms, we find it convenient to express the VARX using compact matrix notation
\begin{align*}
 &   \begin{array}[r]{llll}
       \Y=[\mathbf{y}_1,\dots,\mathbf{y}_T]& (k\times T);  & \quad \utwi{1}=[1,\dots,1]^{\top} & (T \times 1).
\\ 
\mathbf{\Z}_t=
[\mathbf{y}_t^{\top},\dots,\mathbf{y}_{t-p}^{\top},
\mathbf{x}_{t}^{\top},\dots,
\mathbf{x}_{t-s}^{\top} 
]^{\top} & [(kp+ms)\times 1];&\quad
\mathbf{\Z}=
[\bm{\Z}_1,\dots,\bm{\Z}_{T}]
 & [(kp+ms)\times T];\\
\PhiB=[\PhiB^{(1)},\PhiB^{(2)},\dots,\PhiB^{(p)}] & (k\times kp);& \quad \betaB=[\betaB^{(1)},\dots,\betaB^{(s)}]& [k\times ms];\\
  \B=
[\PhiB,\betaB] &
 [k\times (kp+ms)]
;&\quad \mathbf{U}=[\mathbf{u}_1,\dots,\mathbf{u}_T] & (k\times T)
\end{array}
\end{align*}
Equation \eqref{VAR1a} then becomes   
\begin{align*}
 \Y=\utwi{\nu}\utwi{1}^{\top}+\B\mathbf{\Z}+\mathbf{U},
\end{align*}
and the least squares procedure \eqref{eq:ls} can be expressed as
minimizing $\frac{1}{2}\|\Y-\utwi{\nu}\utwi{1}^{\top}-\B\mathbf{\Z}\|_F^2$
over $\utwi{\nu}$ and $\B$.
\subsection{Intercept Term}
\label{secint}

In regularization problems, the intercept $\hat{\utwi{\nu}}$ is not typically regularized and can instead be derived separately.  Using compact matrix notation, we can express the unpenalized portion of (\ref{PenFunForm}) as
\begin{align}
\label{grad2}
f(\B,\utwi{\nu}) &= \frac{1}{2}||\Y-\utwi{\nu}\utwi{1}^{\top}-\B\Z||_F^2,
\end{align}
We can find $\hat{\utwi{\nu}}$ by calculating the gradient of \eqref{grad2} with respect to $\utwi{\nu}$
\begin{align*}
&0=\nabla_{\utwi{\nu}} f(\B,\utwi{\nu})=(\Y-\hat{\utwi{\nu}}\utwi{1}^{\top}-\widehat{\B}\Z)\utwi{1},\\
&\implies \hat{\utwi{\nu}}_k(\lambda)=\widebar{\Y}_{k\cdot}-\widehat{\B}\widebar{\Z}_{k\cdot},  
\end{align*}
in which $\widebar{\Y}_{k\cdot}=\frac{1}{T}\sum_t \Y_{kt},$ and $\widebar{\Z}_{ k\cdot }=\frac{1}{T}\sum_t \Z_{kt}$.  This provides some insight into the scaling, as we can rewrite \eqref{grad2} as
\begin{align}
&\min_{\B}\frac{1}{2}||\Y-(\widebar{\Y}-\B\widebar{\Z})\utwi{1}^{\top}-\B\Z||_F^2,\nonumber\\
\label{Centered}
=&\min_{\B}\frac{1}{2}||(\Y-\widebar{\Y}\utwi{1}^{\top})-\B(\Z-\widebar{\Z}\utwi{1}^{\top})||_F^2,
\end{align}
in which $\widebar{\Y}$ is a $k\times 1$ vector of row means and $\widebar{Z}$ is a $(kp+ms)\times 1$ vector of row means.
\subsection{Solution Strategies}
In the following sections, assume that $\Y$ and $\Z$ are centered as in Equation (\ref{Centered}).
\subsubsection{ Basic VARX-L}
\label{lassovarsol}
Utilizing the coordinate descent framework, we can find $\widehat{\B}$ via scalar updates.  To generalize to a multivariate context, we can express the one-variable update for the $(j,r)$ entry of $\B$, $\B_{jr}$ as
\begin{align}
\label{Onevar}
\min_{\B_{jr}} \frac{1}{2}\sum_t(\Y_{jt}-\sum_{\ell \neq r}\B_{j\ell}\Z_{\ell t}-\B_{jr}\Z_{jt})^2+\lambda|\B_{jr}|.  
\end{align}
Let $\mathbf{R}_t=\Y_{jt}-\sum_{\ell \neq r}\B_{j\ell}\Z_{\ell t}$ denote the partial residual.  Then, we can rewrite Equation (\ref{Onevar}) as
\begin{align*}
&g_{jr}(\B)=\min_{\B_{jr}} \frac{1}{2}\sum_t(\mathbf{R}_t-\B_{jr}\Z_{jt})^2+\lambda|\B_{jr}| \\
=&\min_{\B_{jr}} \frac{1}{2}\sum_{t}(\mathbf{R}_t^2-\B_{jr}^2\Z_{jt}^2-2\mathbf{R}_t\Z_{jt}\B_{jr})+\lambda|\B_{jr}|.
\end{align*}
Now, differentiating with respect to $\B_{jr}$ gives the subgradient as

\begin{align*}
\partial g_{jr}(\B) \ni \B_{jr}\sum_{t}\Z_{jt}^2-\sum_t\mathbf{R}_t\Z_{jt}+\lambda\psi(\B_{jr}),    
\end{align*}
where we define $\psi(\B_{jr})$ as
   \begin{displaymath}
 \psi \in \left\{
     \begin{array}{lr}
       \{\text{sgn}(\B_{jr})\} & \B_{jr}\neq 0 \\
      \left[-1,1\right] & \B_{jr}=0.
     \end{array}
   \right.
\end{displaymath} 
For $\hat{\B}_{jr}$ to be a global minimum, $0\in \partial g(\hat{\B}_{jr})$.
After some algebra, the optimal update can be expressed as
\begin{align*}
 \hat{\B}_{jr} \leftarrow \frac{\mathcal{ST}(\sum_t \mathbf{R}_t\Z_{jt},\lambda)}{\sum_{t}\Z_{jt}^2}. 
\end{align*}
Where $\mathcal{ST}$ represents the soft-threshold operator
\begin{align*}
  \mathcal{ST} (x,\phi)=\sgn(x)(|x|-\phi)_{+},
\end{align*}
$\sgn$ denotes the signum function, and $(|x|-\phi)_+=\max(|x|-\phi,0)$.
The Basic VARX-L procedure is detailed in Algorithm \ref{alg1}.
\subsubsection{Lag Group VARX-L}
\label{grouplassosol}
Rather than vectorizing the Lag Group VARX-L and solving the corresponding univariate least squares problem, if the groups are proper submatrices we can exploit the matrix structure for considerable computational gains.  Without loss of generality, we will consider the ``one lag'' problem for $\PhiB^{(q)}$ (the problem for $\betaB^{(q)}$ is analogous).   
\begin{align}
\label{SP1}
\min_{\PhiB^{(q)}}\frac{1}{2}\|\mathbf{R}_{-q}-\PhiB^{(q)}\Z_q\|_{F}^2+\lambda\|\PhiB^{(q)}\|_F,  
\end{align}
in which, for notational ease, we directly incorporate the weighting into the penalty parameter by defining $\lambda=k\lambda$, $\mathbf{R}_q=\PhiB^{(-q)}\Z_{-q}-\Y \in \mathbb{R}^{k \times T}$ again represents the partial residual.  Taking the gradient of $\|\mathbf{R}_{-q}-\PhiB^{(q)}\Z_q\|_F^2$ with respect to $\PhiB^{(q)}$ results in
\begin{align*}
&\nabla_{\PhiB^{(q)}} \frac{1}{2}\|\mathbf{R}_{-q}-\PhiB^{(q)}\Z_q\|_F^2=\nabla_{\PhiB^{(q)}} \Tr\left((\mathbf{R}_{-q}-\PhiB^{(q)}\Z_q)(\mathbf{R}_{-q}-\PhiB^{(q)}\Z_q)^{\top}\right),\\
=&\PhiB^{(q)}\Z_q\Z_q^{\top}-\mathbf{R}_{-q}\Z_{q}^{\top}.
\end{align*}
In which $\Tr$ denotes the trace operator.
The subgradient of \eqref{SP1} with respect to $\PhiB^{(q)}$ is then
\begin{align*}
\PhiB^{(q)}\Z_q\Z_q^{\top}-\mathbf{R}_{-q}\Z_{q}^{\top}+\lambda\omega(\PhiB^{(q)}),   
\end{align*}
where $\omega$ is defined as
 \begin{displaymath}
   \omega(\PhiB^{(q)}) = \left\{
     \begin{array}{lr}
       \frac{\PhiB^{(q)}}{\|\PhiB^{(q)}\|_{F}} & \PhiB^{(q)}\neq 0\\
       \{U: ||U||_{F}\leq 1\} & \PhiB^{(q)}=0. 
     \end{array}
   \right.
\end{displaymath} 
Consider the case where $\hat{\PhiB}^{(q)}=\bm{0}$.  Then
\begin{align*}
&\frac{\hat{\PhiB}^{(q)}\Z_q\Z_q^{\top}-\mathbf{R}_{-q}\Z_{q}^{\top}}{\lambda}\in \{U : \|U\|_F\leq 1\},\\
&\iff \|\hat{\PhiB}^{(q)}\Z_q\Z_q^{\top}-\mathbf{R}_{-q}\Z_{q}^{\top}\|_F\leq \lambda,\\
&\iff \|\mathbf{R}_{-q}\Z_{q}^{\top}\|_F\leq \lambda.
\end{align*}
We can conclude that $\hat{\PhiB}^{(q)}=0\Longrightarrow \|\mathbf{R}_{-q}\Z_{q}^{\top}\|_F\leq \lambda$.  Now, assuming $\hat{\PhiB}^{(q)}\neq 0$, we have that
\begin{align}
 &\PhiB^{(q)}\Z_q\Z_q^{\top}-\mathbf{R}_{-q}\Z_{q}^{\top}+\lambda(\frac{\PhiB^{(q)}}{\|\PhiB^{(q)}\|_F})=0\nonumber,\\
& \PhiB^{(q)}\Z_q\Z_q^{\top}+\lambda(\frac{\PhiB^{(q)}}{\|\PhiB^{(q)}\|_F})=\mathbf{R}_{-q}\Z_{q}^{\top}\nonumber,\\
\label{mateqn}
&\PhiB^{(q)}\left(\Z_{q}\Z_q^{\top}+\frac{\lambda}{\|\PhiB^{(q)}\|_{F}}\bm{I}_{k}\right)=\mathbf{R}_{-q}\Z_{q}^{\top}.  
\end{align}
Now, since $\Z_{q}\Z_q^{\top}$ is positive semidefinite and $\lambda>0$, we can infer that $\Z_{q}\Z_q^{\top}+\frac{\lambda}{\|\PhiB^{(q)}\|_{F}}\bm{I}_{k}$ is positive definite, hence it is possible to create a trust region subproblem that coincides with Equation (\ref{SP1}).  However, we need to transform $\mathbf{R}_{-q}\Z_{q}^{\top}\in \mathbb{R}^{k\times k}$ into a vector.  Define
\begin{align*}
&\bm{r}_{q}=\text{vec}(\mathbf{R}_{-q}\Z_{q}^{\top}),\\
&\mathbf{G}_{q}=\Z_{q}\Z_q^{\top}\otimes \mathbf{I}_{k},\\
&\bm{\phi}_{q}=\text{vec}(\PhiB^{(q)}),   
\end{align*}
in which $\otimes$ denotes the Kronecker product.  Hence, we can rewrite Equation (\ref{mateqn}) as
\begin{align*}
\bm{\phi}_q^{\top}\left(\mathbf{G}_q+\frac{\lambda}{\|\bm{\phi}_q\|_{F}}\bm{I}_{k^2} \right)=\bm{r}_q.  
\end{align*}
Applying the same transformation to the original subproblem, we can express Equation \ref{SP1} as the trust region subproblem  
\begin{align*}
&\min \frac{1}{2}\bm{\phi}_{q}^{\top}\bm{G}_{q}\bm{G}_q^{\top}\bm{\phi}_q+\bm{r}_q^{\top}\bm{\phi}_q,\\
&\text{s.t.} \|\bm{\phi}_q\|_F\leq \Delta,
\end{align*}
in which $\Delta>0$ denotes to the trust-region radius which corresponds to the optimal solution of Equation \ref{SP1}.  These modifications allow for the use of the block coordinate descent algorithm described in \cite{goldfarb}.  Expanding upon their arguments, by the Karush-Kuhn-Tucker (KKT) conditions, we must have that: $\lambda(\Delta-\|\bm{\phi}_q^{*}\|_F)=0,$ which implies that $\|\bm{\phi}_q^{*}\|_F=\Delta$.  Then, applying Theorem 4.1 of \cite{wright}, we can conclude that
\begin{align}
\label{bstar}
\bm{\phi}_q^{*}=-\left( \bm{G}_{q}+\frac{\lambda}{\Delta}\bm{I}_{k^2}\right)^{-1}\bm{r}_{q}.
\end{align}
\cite{goldfarb} remarks that Equation (\ref{bstar}) can also be expressed as $\bm{\phi}_q^{*}=\Delta y_q(\Delta)$, where
\begin{align}
\label{eigen1}
 y_q(\Delta)=-\left(\Delta \bm{G}_q+\lambda\bm{I}_{k^2} \right)^{-1}\bm{r}_{q}, 
\end{align}
Note that, based on the KKT conditions, $\|y_q(\Delta)\|_F=1$.
Hence, the optimal $\Delta$ can be chosen to satisfy $\|y_q(\Delta)\|_F=1$.  We can efficiently compute $\|y_q(\Delta)\|_F^2$ via an eigen-decomposition of $\bm{G}_q$.  We start by rewriting Equation \eqref{eigen1} as
\begin{align*}
 &y_q(\Delta)=-\left(\Delta \bm{W} V \bm{W}^{\top}+\lambda\bm{I}_{k^2} \right)^{-1}\bm{r}_{q},\\
=&- \bm{W}(\Delta V+\lambda\bm{I}_{k^2})^{-1}\bm{W}^{\top}\bm{r}_{q},
\end{align*}
in which the first line follows from the spectral decomposition of a symmetric positive semidefinite matrix.  
Finally, we can express $\|y_q(\Delta)\|_F^2$ as 
\begin{align*}
\|y_q(\Delta)\|_F^2=\sum_i \frac{(\mathbf{w_i}^{\top}\bm{r}_q)^2}{(\mathbf{v}_i\Delta+\lambda)^2},  
\end{align*}
in which $\bm{w}_i$ denotes the columns of $\bm{W}$ and $\bm{v}_i$ the diagonal elements of $\bm{V}$.  \cite{goldfarb} notes that we can determine the optimal $\Delta$ by applying Newton's method to find the root of
\begin{align}
\label{NEWTON}
\Omega(\Delta)=1-\frac{1}{\|y_q(\Delta)\|_F}.  
 \end{align}
The full Lag Group VARX-L procedure is detailed in Algorithm \ref{algGL}.  Our algorithm organizes iterations around an ``active-set'' as described in \cite{friedman}.  This approach starts by cycling through every group and then only iterating on the subset of $\B$ that are nonzero (the ``active-set'') until convergence.  If a full pass through all $\B$ does not change the active set, the algorithm has converged, otherwise the process is repeated.  This approach considerably reduces computation time, especially for large values of $\lambda$ in which most model coefficients are zero.   
\subsubsection{Own/Other Group VARX-L}
\label{grouplassooosol}
In the Own/Other setting since the groups are not proper submatrices, in order to properly partition each $\PhiB^{(\ell)}$ into separate groups for own and other lags, Equation \eqref{PenFunForm} must be transformed into a least squares problem.
To perform a least squares transformation, we define the following 
\begin{align*}
&r_{-qq}=\text{vec}(\mathbf{R}_{-qq}),\\
&\bm{\phi}_{qq}=\text{vec}(\PhiB_{\text{on}}^{(q)}),\\
&\bm{M}_{qq}=(\Z^{\top}\otimes I_k)_{qq}.
\end{align*}
Then, the one block subproblem for own lags (group qq) can be expressed as
\begin{align*}
&\min_{\bm{\phi}_{qq}} \frac{1}{2}\|\bm{M}_{qq}\bm{\phi}_{qq}+r_{-qq}\|_F^2+\lambda\|\bm{\phi}_{qq}\|_F,\\
=&\min_{\bm{\phi}_{qq}} \frac{1}{2} r_{-qq}^{\top}r_{qq}+\bm{\phi}_{qq}^{\top}\bm{M}_{qq}^{\top}\bm{M}_{qq}\bm{\phi}_{qq}+r_{-qq}^{\top}\bm{M}_{qq}\bm{\phi}_{qq}+\lambda\|\bm{\phi}_{qq}\|_F, \\
=&\min_{\bm{\phi}_{qq}} \frac{1}{2} \bm{\phi}_{qq}^{\top}\bm{M}_{qq}^{\top}\bm{M}_{qq}\bm{\phi}_{qq}+r_{-qq}^{\top}\bm{M}_{qq}\bm{\phi}_{qq}+\lambda\|\bm{\phi}_{qq}\|_F.
\end{align*}
At $\hat{\bm{\phi}}_{qq}$, we must have that $0\in \partial f(\hat{\bm{\phi}}_{qq})$.  The subgradient can be expressed as
\begin{align*}
\frac{\partial}{\partial \bm{\phi}_{qq}}=\bm{M}_{qq}^{\top}\bm{M}_{qq}\bm{\phi}_{qq}+\bm{M}_{qq}^{\top}r_{qq}+\lambda \omega(\bm{\phi}_{qq}), 
\end{align*}
where $\omega$ is defined as
   \begin{displaymath}
 \omega(s) \in \left\{
     \begin{array}{lr}
       \{\frac{s}{\|s\|_F}\} & s\neq 0 \\
      \{u : \|u\|_F\leq 1\} & s=0.
     \end{array}
   \right.
\end{displaymath} 
Thus, after applying these transformations, we can apply a slightly adapted version of Algorithm \ref{algGL}.
\subsubsection{Sparse Lag Group VARX-L}
\label{sglsol}
As with the Lag Group VARX-L, we will consider the one-block subproblem for lag $\PhiB^{(q)}$
\begin{align}
\label{SGL2}
\min_{\PhiB^{(q)}} \frac{1}{2k}\|\mathbf{R}_{-q}-\PhiB^{(q)}\Z_q\|_{F}^2+(1-\alpha)\lambda\|\PhiB^{(q)}\|_{F}+\alpha\lambda\|\PhiB^{(q)}\|_{1}. 
\end{align}
Since the inclusion of within-group sparsity does not allow for separability, coordinate descent based procedures are no longer appropriate, therefore, following \cite{simon} our solution to the Sparse Lag Group VARX-L utilizes gradient descent based methods.  We express Equation \eqref{SGL2} as the sum of a generic differentiable function with a Lipschitz gradient and a non-differentiable function.  

We start by linearizing the quadratic approximation of the unpenalized loss function that only makes use of first-order information around its current estimate $\PhiB_{0}$ (borrowing from \cite{simon}, for notational ease, let $\bm{\PhiB}\equiv \bm{\PhiB}^{(q)}$, $\ell(\bm{\PhiB})$ represent the unpenalized loss function, and $\mathcal{P}(\bm{\PhiB})$ represent the penalty term).  Then, we can express the linearization as
\begin{align*}
&M(\bm{\PhiB},\PhiB_{0})=\ell(\PhiB_{0})+\text{vec}(\bm{\PhiB}-\PhiB_{0})^{\top}\text{vec}(\nabla\ell(\PhiB_{0}))+\frac{1}{2d}\|\PhiB-\PhiB_{0}\|_{F}^2+\mathcal{P}(\PhiB)\\
=&\frac{1}{2k}\|\mathbf{R}_{-q}-\PhiB_{0}\Z_{q}\|_{F}^2+\langle \PhiB-\PhiB_{0},(\PhiB_{0}\Z_q-\mathbf{R}_{-q})\Z_{q}^{\top}\rangle +\frac{1}{2d}\|\PhiB-\PhiB_{0}\|_{F}^2+\mathcal{P}(\PhiB),
\end{align*}
in which $d$ represents the step size.  Removing terms independent of $\PhiB$, our objective function becomes
\begin{align*}
&\argmin_{\PhiB} M(\bm{\PhiB},\PhiB_{0}),\\
=&\argmin_{\bm{\PhiB}} \frac{1}{2d}\|\bm{\PhiB}-\left( \PhiB_{0}-d(\PhiB_{0}\Z_q-\mathbf{R}_{-q})\Z_{q}^{\top}\right)\|_{F}^2+\mathcal{P}(\PhiB).
\end{align*}
Then, generalizing the arguments outlined by \cite{simon}, we can infer that the optimal update $U(\PhiB)$ can be expressed as
\begin{align*}
U(\PhiB)=\left( 1-\frac{d(1-\alpha)\lambda}{\|ST(\PhiB_{0}-d(\PhiB_{0}\Z_{q}-\mathbf{R}_{-q})\Z_{q}^{\top},d\alpha\lambda)\|_{F}}\right)_{+}ST(\PhiB_{0}-d(\PhiB_{0}\Z_{q}-\mathbf{R}_{-q})\Z_{-q}^{\top},d\alpha\lambda). 
\end{align*}
As in \cite{simon}, we apply a Nesterov accelerated update.  At step j, we update according to
\begin{align}
\label{NESTA}
\hat{\PhiB}[j] \leftarrow \hat{\PhiB}[j-1]+\frac{j}{j+3}(U(\PhiB)-\hat{\PhiB}[j-1]),   
\end{align}
which, per \cite{beck} converges at rate $1/j^2$ as opposed to the $1/j$ rate of the standard proximal gradient descent.

We use a constant step size according to the Lipschitz constant, H, which must satisfy
\begin{align*}
||\nabla_X \ell(X)-\nabla_Y \ell(Y)||\leq H ||X-Y||.
\end{align*}
Consider two submatrices $\mathbf{A}^{(q)}$ and  $\mathbf{C}^{(q)}$.  We have that
\begin{align*}
&\nabla_{\mathbf{A}^{(q)}} \ell(\mathbf{A}^{(q)})=\mathbf{A}^{(q)}\Z_q\Z_q^{\top}-\mathbf{R}_{-q}\Z_q^{\top},\\
& \nabla_{\mathbf{C}^{(q)}} \ell(\mathbf{C}^{(q)})=\mathbf{C}^{(q)}\Z_q\Z_q^{\top}-\mathbf{R}_{-q}\Z_q^{\top},\\
\implies &\nabla_{\mathbf{A}^{(q)}} \ell(\mathbf{A}^{(q)}) - \nabla_{\mathbf{C}^{(q)}} \ell(\mathbf{C}^{(q)})=(\mathbf{A}^{(q)}-\mathbf{C}^{(q)})\Z_q\Z_q^{\top},\\
\implies & \|(\mathbf{A}^{(q)}-\mathbf{C}^{(q)})\Z_q\Z_q^{\top}\|_2\leq \|\mathbf{A}^{(q)}-\mathbf{C}^{(q)}\|_2\|\Z_q\Z_q^{\top}\|_2.
\end{align*}
The last inequality follows from the sub-multiplicity of the matrix 2-norm.  Therefore, we can conclude that the Lipschitz constant is $||\Z_q\Z_q^{\top}||_2=\sigma_{1}(\Z_q\Z_q^{\top})$, i.e. the largest singular value of $\Z_q\Z_q^{\top}$, which has dimension $k\times k$ for $\PhiB^{(1)},\dots,\PhiB^{(p)}$ and is a scalar for exogenous groups.  Since $\Z_q\Z_q^{\top}$ is symmetric and positive semidefinite, it is diagonalizable, and the maximum eigenvalue can be efficiently computed using the power method, described in \cite{Golub}.

As only the maximum eigenvalue is required, the power method is much more computationally efficient than a computation of the entire eigensystem.  Moreover, we retain the corresponding eigenvector produced by this procedure to use as a ``warm start'' that substantially decreases the amount of time required to compute the maximal eigenvalue at each point in time during the cross-validation and forecast evaluation procedures.

In a manner similar to Algorithm \ref{algGL}, an ``active-set'' approach is used to minimize computation time.  The inner loop of  of the Sparse Group VARX-L procedure is detailed in Algorithm (\ref{algSGL}).  An outline of the algorithm is below:
\begin{enumerate}
\item Iterate through all groups.  For each group:
  \begin{enumerate}
  \item Check if the group is active via the condition: $||(\PhiB^{(q)}\Z_{q}-\mathbf{R}_{-q})\Z_{q}^{\top}||_{F}\leq (1-\alpha)\lambda$. 
  \item If active, go to the inner loop (Algorithm \ref{algSGL}), if not active, set group identically to zero.
  \item Repeat until convergence.
  \end{enumerate}
\end{enumerate}

Upon performing the least squares transformations as in the Own/Other Group VARX-L (section \ref{grouplassooosol}), the Sparse Own/Other Group VARX-L follows almost the exact same procedure as its lag counterpart.

\subsubsection{Endogenous-First VARX-L}
\label{HVARX1}
The Endogenous-First VARX-L is of the form 
\begin{align*}
\min_{\PhiB,\betaB} \frac{1}{2}\|\Y-\B\Z\|_F^2+\lambda \sum_{\ell=1}^p\sum_{j=1}^k\bigg(\|[\PhiB_j^{(\ell)},\betaB_{j,\cdot}^{(\ell)}]\|_F+\|\betaB_{j,\cdot}^{(\ell)}\|_F\bigg).
\end{align*}
Since the optimization problem decouples across rows, we will consider solving the \emph{one row} subproblem (for row i)
\begin{align}
\label{HV1}
\min_{\PhiB_i,\betaB_i} \frac{1}{2}\|\Y_i-\B_i\Z\|_F^2+\lambda \sum_{\ell=1}^p\bigg(\|[\PhiB_i^{(\ell)},\betaB_{i,\cdot}^{(\ell)}]\|_F+\|\betaB_{i,\cdot}^{(\ell)}\|_F\bigg).
\end{align}
In a manner similar to the Sparse Group VARX-L, the Endogenous-First VARX-L is solved via proximal gradient descent.  For ease of notation, let $\mathcal{P}(\PhiB,\betaB)$ represent the nested penalty.  The update step for the Endogenous-First VARX-L (at step $j$) can be expressed as
\begin{align}
\label{XUS}
\B_i[j]=\text{Prox}_{d\lambda,\mathcal{P}(\PhiB,\betaB)}\big(\B_i[j-1]-d\nabla \ell(\B_i[j-1])\big), 
\end{align}
in which $d$ denotes step size and $\ell(\B_i[j-1])$ denotes the unpenalized loss function.  Note that $\nabla \ell (\B_i)=-(\Y_i-\B_i\Z)\Z^{\top}$.  Similar to the Sparse Group VARX-L setting, a fixed step size is used; $d=\frac{1}{\sigma_1(\Z\Z^{\top})}$.
To speed convergence, as in the Sparse Group VARX-L update step \eqref{NESTA}, we apply a similar Nesterov-style accelerated update: 
\begin{align*}
\hat{\B}_i \leftarrow \hat{\B}_i[j-1]+\frac{j-2}{j+1}(\hat{\B}_i[j-1]-\hat{\B}_i[j- 2]),     
\end{align*}
Thus, \eqref{XUS} becomes 
\begin{align}
\label{XUS2}
\B_i[j]=\text{Prox}_{d\lambda,\mathcal{P}(\PhiB,\betaB)}(\hat{\B}_i -d\nabla \ell(\hat{\B}_i) ), 
\end{align}
 \cite{Jenatton} observed that the dual of \eqref{XUS2} can be solved with one pass of block coordinate descent.  Moreover, the block updates are extremely simple and available in closed-form.  Algorithm \ref{ALGHV} details the prox function for the Endogenous-First VARX-L.  Note that it consists of $p$ separate nested structures for each series.  Thus, solving \eqref{XUS2} essentially amounts to calling the same proximal function $p$ times at each update step.
\subsection{\cite{BGR} Implementation}
\label{BGRAPP}
The Bayesian VAR proposed by \cite{BGR} utilizes a normal inverted Wishart Prior.  Defining $\phi=\text{vec}(\PhiB)$, the prior has the form
\begin{align*}
&\phi|\Omega \sim N(\phi_0,\Omega \otimes \Omega_0)\\
&\Omega \sim iW(S_0,\alpha_0),  
\end{align*}
in which iW denotes the inverse Wishart distribution.  This prior is implemented by adding the following dummy observations to $\Y$ and $\Z^{\top}$ (which we define as $\X$):
\begin{align*}
 &\Y_{d_1}=
 \begin{pmatrix}
   \text{diag}(\delta\sigma_1,\dots,\delta\sigma_k)/\lambda\\
   \boldmath{0}_{k \times (p-1) \times k}\\
   \text{diag}(\sigma_1,\dots,\sigma_k)\\
  \boldmath{0}_{1\times k}
 \end{pmatrix}
\\
&\X_{d_1}=
\begin{pmatrix}
\boldmath{0}_{kp\times 1} & J_p\otimes \text{diag}(\sigma_1,\dots,\sigma_k)\\
\boldmath{0}_{k\times 1} & \boldmath{0}_{k\times kp}\\
\epsilon   &  \boldmath{0}_{1\times kp}
\end{pmatrix}.
\end{align*}

The scale parameters for the prior variances of each series, $\sigma_1,\dots,\sigma_k$, are estimated by univariate autoregressive models.  $J_p=\text{diag}(1,2,\dots,p)$, and $\epsilon$ denotes the prior on  the intercept and is set to a very small number (e.g. 1e-5).  $\delta$ serves as an indicator for the prior belief that the series have high persistence.  We set $\delta=0$ in all of our forecasting applications except for the Minnesota VARX-L application in section $\ref{sec8}$.  

In addition to the above construction, following \cite{Doan}, BGR adds a prior that imposes a bound on the sum of coefficients by shrinking $\Pi=(I_k-\PhiB_1-\dots-\PhiB_p)$ toward zero.  This prior is implemented by adding the additional dummy observations 
\begin{align*}
&\Y_{d_2}=\text{diag}(\delta\mu_1,\dots,\delta\mu_k)/\tau, \\
&\X_{d_2}=
\begin{pmatrix}
 \utwi{0}_{k\times 1} & 1_{1\times p}\otimes \text{diag}(\delta\mu_1,\dots,\delta\mu_k)/\tau    
\end{pmatrix},
\end{align*}
in which $\mu_1,\dots,\mu_k$ are meant to capture the average level of each of the series, set according to their unconditional means and $\tau$ denotes a loose prior which is set to $10\lambda$.  After appending the dummy observations to $\Y$ and $\X$ and creating the augmented matrices $\Y_{*}$ and $\X_{*}$, the posterior mean can be calculated in closed form as:
\begin{align*}
\tilde{\PhiB}^{\top}=(\X_{*}^{\top}\X_{*})^{-1}\X_{*}^{\top}\Y_{*}  
\end{align*}

\subsection{Penalty Grid Selection}
\label{pengrid}
  \begin{table}[H]
\footnotesize
    \centering    
    \caption{\label{tab:tabSP} \footnotesize Starting values of the penalty grid for each procedure; $\rho_q$ denotes the number of variables in group q.
} 
   \begin{tabular}{ l | l  }
Structure& Starting Value of $\Lambda_{\text{Grid}}$ \\
    \hline
\rule{0pt}{3ex}    
Basic  & $\|\Z\Y^{\top}\|_{\infty}$ \\
 Lag Group  &  $\max_q (\|\Z_q\Y^{\top}\|_F)$   \\
 Sparse Lag  &  $\max_q \big(\|\Z_q\Y^{\top}\|_F\big)\alpha$   \\
Own/Other Group  & $\max_q\big(\|(\Z^{\top}\otimes \utwi{I}_{k})_q \text{vec}(\Y)\|_F/\sqrt{\rho_q}\big)$\\
 Sparse Own/Other Group  & $\max_q\big(\|(\Z^{\top}\otimes \utwi{I}_{k})_q \text{vec}(\Y)\|_F/\sqrt{\rho_q}\big)\alpha$\\
Endogenous-First & $\max_i (\|\Z\Y_i^{\top}\|_F)$,
  \end{tabular}
 \end{table}
\subsection{Algorithms}
\label{algs}
\begin{algorithm}
\footnotesize
\caption{$\text{Basic-VARX-L}_{k,m}(p,s)$}
\label{alg1}
\begin{algorithmic}[5]
\Require $\Y,\Z,\B^{\text{INI}},\lambda$   
\State $\B^{\text{OLD}}$ $\leftarrow$  $\B^{\text{INI}}$
\Repeat
 \For{$i =1,\dots, k$,  $j=1,\dots,kp+ms$}
\State  $\utwi{R} \leftarrow \Y_{i\cdot}-\sum_{\ell  \neq j}\B_{i\ell}\Z_{\ell \cdot}$ 
\State $\B_{ij}^{\text{NEW}}$ $\leftarrow$ $ \frac{\text{ST}(\sum_t \utwi{R} \Z_{j\cdot},\lambda)}{\sum_{t}\Z_{jt}^2}$
\EndFor
\State $\B^{\text{OLD}}\leftarrow \B^{\text{NEW}}$
\Until Desired threshold is reached
\State $\hat{\utwi{\nu}} \leftarrow \bar{\Y}-\B^{\text{NEW}}\bar{\Z}$\\
\Return $\hat{\utwi{\nu}},\B^{\text{NEW}}$
\end{algorithmic}
\end{algorithm}
\begin{algorithm}
\footnotesize
\caption{\label{alg2} Basic VARX-L(p,s) Cross-Validation}
\begin{algorithmic}[5]
\Require{$\Y,\Z,\B^{\text{INI}},\Lambda_{\text{grid}}$,$h$}  
\State $\B^{\text{LAST}}\leftarrow \B^{\text{INI}}$
\State \For{t  in $[T_1,T_2-h]$}
\State $\Y_{\text{TRAIN}}^{(t)}$ $\leftarrow$ $\Y_{,h:(t-1)}$
\State $\Z_{\text{TRAIN}}^{(t)}$ $\leftarrow$ $\Z_{,1:(t-h+1)}$
\For {i in $\Lambda_{\text{Grid}}$}
\State $\utwi{\nu}_i,\B_i^{\text{NEW}}\leftarrow$ \text{Basic-VARX-L}($\Y_{\text{TRAIN}}^{(t)},\Z_{\text{TRAIN}}^{(t)},\B_{i}^{\text{LAST}},\lambda_{i})$
\State $\Z_{\text{TEST}}^{(t)}\leftarrow \Z_{,(t+1)}$
\State $SSFE^{(t,i)}$ $\leftarrow$ $\|\Y_{t+h}-[\nu_i,\B_i^{\text{NEW}}]*[\utwi{1},\Z_{TEST}^{(t)}]\|_F^2$
\State $\B_{i}^{\text{LAST}}\leftarrow \B_{i}^{\text{NEW}}$\\ 
\EndFor
\For {i in $\Lambda_{\text{Grid}}$}
\State $MSFE^{(i)} \leftarrow \frac{1}{T_2-T_1-h+1}\sum_t SSFE^{(t,i)}$
\EndFor\\
\EndFor\\
\Return $\lambda_{\hat{i}}$, where $\hat{i}=\text{argmin}_i$ $MSFE^{(i)}$
\end{algorithmic}
\end{algorithm}

\begin{algorithm}
\footnotesize
\caption{$\text{Lag Group VARX-L}_{k,m}(p,s) \text{ with active-set strategy}$}
\label{algGL}
\begin{algorithmic}[5]
\Require $\B_{\text{INI}},\mathcal{G},\Y,\Z, \mathcal{A}_{\text{INI}},\Lambda$
\State Define:
\begin{align*}
\text{ for } g=1,\dots,p+ms:\\
&\mathbf{G}_{g}=\mathbf{M}_{g}\otimes \mathbf{I}_{k}.
\end{align*}
\State $\B_{\lambda_0,\text{INI}} \leftarrow \B_{\text{INI}}$, 
\State $\mathcal{A}_{\lambda_0,\text{INI}} \leftarrow \mathcal{A}_{\text{INI}}$, 
\For{ $\lambda \in \Lambda_{\text{Grid}}$}
\Repeat
\State $\B_{\lambda,\mathcal{A}_{\lambda}} \leftarrow $ThresholdUpdate($\mathcal{A}_{\lambda},\B_{\lambda,\mathcal{A}_{\lambda}},\lambda$) 
\State $\B_{\lambda,\mathcal{A}_{\text{FULL}}}, \mathcal{A}_{\lambda} \leftarrow$BlockUpdate($\mathcal{A}_{\text{FULL}},\B_{\lambda,\mathcal{A}_{\lambda}},\lambda$)
\Until $\B_{\lambda,\mathcal{A}_{\lambda}}=\B_{\lambda,\mathcal{A}_{\text{FULL}}}$
\State $\hat{\nu} \leftarrow \bar{\Y}-\B_{\lambda,\mathcal{A}}\bar{\Z}$
\EndFor\\
\Return $\hat{\utwi{\nu}}, \B_{\Lambda}, A_{\Lambda}$
\Procedure {BlockUpdate}{$\mathcal{G},\B_{\text{INI}},\lambda$}\Comment{Makes one full pass through all groups}
\State $\B\leftarrow \B_{\text{INI}}$
\For {$g\in \mathcal{G}$}
\State $\mathbf{R} \leftarrow \B_{-g}\Z_{-g}-\Y$
\State $\mathbf{r}\leftarrow \mathbf{R}\Z_{q}^{\top}$
\If {$\|\mathbf{r}\|_F \leq \lambda$}
\State $\B_g^{*}\leftarrow \utwi{0}_{|g|}$
\State $\mathcal{A}_{g}\leftarrow \emptyset$ 
\EndIf
\If {$\|\mathbf{r}\|_F > \lambda$}
\State $\Delta \leftarrow$ the root of $\Omega(\Delta)$ defined in (\ref{NEWTON})
\State $\text{vec}(\B_{g})\leftarrow -(\mathbf{G}_{g}+\frac{\lambda}{\Delta}\mathbf{I})^{-1}\mathbf{r}$
\State $\mathcal{A}_{g}\leftarrow g$
\EndIf
\EndFor\\
\Return $\B_{\lambda}$, $\mathcal{A}$
\EndProcedure
\Procedure {ThresholdUpdate}{$\mathcal{A}_{\lambda},\B_{\lambda,\text{INI}},\lambda$} \Comment{Iterates through active set until convergence}
\If {$\mathcal{A}_{\lambda}=\emptyset$}
\Return $\utwi{0}_{k\times kp+ms}$
\EndIf
\If {$\mathcal{A}_{\lambda}\neq \emptyset$}
\State $\B_{\lambda,\text{OLD}}\leftarrow \B_{\lambda,\text{INI}}$
\Repeat
\State $\B_{\lambda,\text{NEW}}, \mathcal{A}_{\lambda} \leftarrow$ BlockUpdate($\mathcal{A}_{\lambda},\B_{\lambda,\text{OLD}},\lambda$)
\State $\B_{\lambda,\text{OLD}}\leftarrow \B_{\lambda,\text{NEW}}$
\Until Desired threshold is reached
\EndIf \\
\Return $\B_{\lambda,\text{NEW}}, \mathcal{A}$ 
\EndProcedure
\end{algorithmic}
\end{algorithm}
\begin{algorithm}
\footnotesize
\caption{Sparse Lag Group VARX-L inner loop}
\label{algSGL}
\begin{algorithmic}[5]
\Require $\PhiB_0,\Z_{q},\mathbf{R}_{-q},\alpha$
\State $h\leftarrow \frac{1}{\sigma_1(\Z_{q})^2}$
\State $\PhiB_0\leftarrow \PhiB^{1}$
\Repeat
\State $j$ $\leftarrow$1
\State $\bm{F}_{q} \leftarrow \frac{(\PhiB^{j}\Z_{q}-\mathbf{R}_{-q})\Z_q^{\top}}{k}$
\State $\gamma^{j}\leftarrow \PhiB^{j}$
 \State $\text{vec}(\bm{\gamma}^{(j+1)})\leftarrow\left(1-\frac{h(1-\alpha)\lambda}{\|ST(\PhiB^{j}-h(\PhiB^{j}\Z_{q}-\mathbf{R}_{-q})\Z_{q}^{\top},h\alpha\lambda)\|_{F}}\right)_{+}ST(\text{vec}(\PhiB^{j})-h\text{vec}(\bm{F}_q),h\alpha\lambda) $\\
\State $\PhiB^{j+1}\leftarrow \gamma^{j+1}+\frac{j}{j+3}(\gamma^{j+1}-\gamma^{j})$
\State $j\leftarrow j+1$
\Until Desired threshold is reached
\end{algorithmic}
\end{algorithm}
\setlength{\floatsep}{1pt}
\begin{algorithm}[!h]
\caption{\label{ALGHV} Endogenous-First VARX-L Proximal Problem  }
\begin{algorithmic}
\Require $\tilde v,\lambda,k,p,m,s $  
\For{i=1,\dots,p}
\State $g_1 \leftarrow [\big((i-1)\cdot k+1\big):\big((i-1)\cdot k+k\big)]$
\State $g_2 \leftarrow [\big( (i-1)\cdot m+p\cdot k\big):\big((i-1)\cdot m+p\cdot k+m\big)]$ 
\State $v_{\{g_1,g_2\}}\leftarrow \text{Prox}(v_{\{g_1,g_2\}},\lambda,k)$
\EndFor\\
\Return $ v$.
\Procedure {Prox}{$v$,$\lambda$,k}
\State $h_2\leftarrow (k+1):(k+m)$
\State $h_1\leftarrow 1:(k+m)$
\For{$j=1,2$}
\State $v_{h_j}\leftarrow (1 -\lambda/\|v_{h_j}\|_F)_+v_{h_j}$
\EndFor\\
\Return $ v$.
\EndProcedure

\end{algorithmic}
\end{algorithm}

\end{document}